\documentclass[useAMS,usenatbib,usegraphicx]{mn2e}

\title[ULTRACAM: an ultra-fast, triple-beam CCD camera]{ULTRACAM: an 
ultra-fast, triple-beam CCD camera for high-speed astrophysics}

\author[V. S. Dhillon et al.]
{V. S. Dhillon,$^1$\thanks{E-mail: vik.dhillon@shef.ac.uk}
T. R. Marsh,$^2$\thanks{E-mail: t.marsh@warwick.ac.uk}
M. J. Stevenson,$^1$ 
D. C. Atkinson,$^3$
P. Kerry,$^1$ 
\newauthor
P. T. Peacocke,$^3$ 
A. J. A. Vick,$^3$ 
S. M. Beard,$^3$
D. J. Ives,$^3$
D. W. Lunney,$^3$ 
\newauthor
S. A. McLay,$^3$ 
C. J. Tierney,$^3$
J. Kelly,$^1$ 
S. P. Littlefair,$^1$
R. Nicholson,$^1$ 
\newauthor
R. Pashley,$^1$ 
E. T. Harlaftis,$^4$
K. O'Brien$^5$
\\
$^{1}$Department of Physics and Astronomy, University of Sheffield, 
Sheffield S3 7RH, UK \\
$^{2}$Department of Physics, University of Warwick, Coventry CV4 7AL, UK \\
$^{3}$UK Astronomy Technology Centre, Royal Observatory Edinburgh, 
Blackford Hill, Edinburgh EH9 3HJ, UK \\
$^{4}$Institute of Space Applications and Remote Sensing, National Observatory of Athens, I. Metaxa and B. Paulou, \\
$^{\ }$Lofos Koufou, Palaia Penteli, Athens 152 36, Greece \\
$^{5}$European Southern Observatory, Alonso de Cordova 3107, Vitacura, Casilla 19001, Santiago 19, Chile
}

\begin{document}

\date{Accepted. Received; in original form}

\pagerange{\pageref{firstpage}--\pageref{lastpage}} \pubyear{2006}

\maketitle

\label{firstpage}

\begin{abstract}

  ULTRACAM is a portable, high-speed imaging photometer designed to
  study faint astronomical objects at high temporal
  resolutions. ULTRACAM employs two dichroic beamsplitters and three
  frame-transfer CCD cameras to provide three-colour optical imaging
  at frame rates of up to 500 Hz. The instrument has been mounted on
  both the 4.2-m William Herschel Telescope on La Palma and the 8.2-m
  Very Large Telescope in Chile, and has been used to study white
  dwarfs, brown dwarfs, pulsars, black-hole/neutron-star X-ray
  binaries, gamma-ray bursts, cataclysmic variables, eclipsing binary
  stars, extrasolar planets, flare stars, ultra-compact binaries,
  active galactic nuclei, asteroseismology and occultations by Solar
  System objects (Titan, Pluto and Kuiper Belt objects). In this paper
  we describe the scientific motivation behind ULTRACAM, present an
  outline of its design and report on its measured performance.

\end{abstract}


\section{Introduction}
\label{sec:introduction}

Charge-Coupled Devices (CCDs) revolutionised astronomy when they
became available in the 1970's. CCDs are linear, stable, robust and
low-power devices. They have large formats, small pixels and excellent
sensitivity over a wide range of wavelengths and light levels; in
fact, they are almost perfect detectors, suffering only from poor time
resolution and readout noise compared to the photon-counting detectors
that they replaced.  These limitations of CCDs are inherent to their
architecture, in which the photo-generated electrons must first be
extracted (or {\em clocked}) from the detection site and then
digitised.\footnote{For the purposes of this paper, the word {\em
digitisation} shall refer to both the process of determining the
charge content of a pixel via correlated double sampling and the subsequent
digitisation of the charge using an analogue-to-digital converter.}

There are ways in which the readout noise of CCDs can be eliminated --
e.g. by the use of electron-multiplying devices \citep{mackay01}.
There are also ways in which CCDs can be made to read out
faster. First, the clocking rate can be increased and the digitisation
time decreased, but only if the CCD is of sufficient quality to allow
these adjustments to be made without an unacceptable increase in
charge-transfer inefficiency and readout noise.  Second, the duty
cycle of the exposures can be increased; for example, {\em
  frame-transfer} CCDs provide a storage area into which
photo-generated charge can be clocked. This charge is then digitised
whilst the next exposure is taking place and, because digitisation
generally takes much longer than clocking, the dead time between
exposures is significantly reduced. Third, the data acquisition
hardware and software can be designed in such a way that the rate at
which it is able to archive the data (e.g. to a hard disk) is always
greater than the rate at which data is digitised by the CCD -- a
situation we shall refer to as {\em detector-limited} operation.

We have employed all three of the techniques described above to
harness the greater sensitivity and versatility of CCDs (compared to
photon-counting detectors) for high-speed optical photometry. The
resulting instrument, known as ULTRACAM (for ULTRA-fast CAMera), was
commissioned on the 4.2-m William Herschel Telescope (WHT) on La Palma
on 16 May 2002, and on the 8.2-m Very Large Telescope (VLT) in Chile
on 4 May 2005. To date, no detailed description of the instrument has
appeared in the refereed astronomical literature. In this paper,
therefore, we describe the scientific motivation behind ULTRACAM,
present an outline of its design and report on its measured
performance.

\section{Scientific motivation}
\label{sec:requirements}

It is widely recognised that temporal resolution is a relatively
unexplored region of observational parameter space (see, for example,
\citealt{dravins94}), particularly in the optical part of the
spectrum. There is a simple reason for this: CCDs are the dominant
detector on all of the largest ground-based telescopes and the
resources of the observatories have understandably been targeted
towards increasing the area and sensitivity of their detectors rather
than prioritising high-speed readout. As a result, it has been
difficult for astronomers to achieve frame rates of higher than $\sim
1$ frame per minute using CCDs on large optical telescopes, although
there are a few notable exceptions (see table~1 for details).

\begin{table*}
\centering
\caption{Instruments/modes for high-speed astrophysics with CCDs --
  this is not intended to be an exhaustive list, but an indication of
  the variety of ways in which CCDs can be used for high-speed optical
  observations. Those listed with a dagger ($^{\dag}$) symbol are not
  continuous modes and data-taking has to be periodically stopped for
  archiving. The frame rates listed are estimates of the maximum values
  possible with each mode. Abbreviations: F-T -- frame-transfer; MCP --
  micro-channel plate; EMCCD -- electron-multiplying CCD.}
\begin{tabular}{lllll}
\hline\noalign{\smallskip}
Instrument/mode & Telescope & Detector & Frame rate & Reference \\
\hline\noalign{\smallskip}
Low-smear drift (LSD) mode$^{\dag}$      & WHT          & CCD     & 4 Hz     & \cite{rutten97a} \\
Time-series mode$^{\dag}$                & AAT          & CCD     & 100 Hz   & \cite{stathakis02} \\
Phase-binning mode                       & Hale         & CCD     & 1000 Hz  & \cite{kern02a}, \cite{kern02b} \\
Stroboscopic mode                        & GHO          & CCD     & 7.5 Hz   & \cite{cadez01}, \cite{kotar03} \\
Freerun mode                             & Mayall       & CCD+MCP & 24000 Hz & \cite{fordham00}, \cite{fordham02} \\
Continuous-clocking mode                 & Keck-II      & CCD     & 14 Hz    & \cite{obrien01b}; \cite{skidmore03} \\
High time-resolution (HIT) mode$^{\dag}$ & VLT          & CCD     & 833 Hz   & \cite{cumani01}, \cite{obrien07} \\
ULTRACAM                                & WHT/VLT      & CCD     & 500 Hz   & This paper \\
Frame-storage mode                      & Lick 1-m     & F-T CCD     & 0.1 Hz     & \cite{stover87} \\
UCT photometer                           & SAAO         & F-T CCD & 1 Hz     & \cite{odonoghue95}, \cite{woudt01} \\
Acquisition camera                       & Gemini South & F-T CCD & 7 Hz     & \cite{hynes03} \\
JOSE camera$^{\dag}$                     & NOT          & F-T CCD & 150 Hz   & \cite{stjacques97}, \cite{baldwin01} \\
SALTICAM                                 & SALT        & F-T CCD   & 10 Hz  & \cite{odonoghue06} \\
LuckyCam                                 & NOT   & EMCCD   & 30 Hz  & \cite{law06} \\
\hline\noalign{\smallskip}
\end{tabular}
\label{tab:competitors}
\end{table*}

Aside from the serendipitous value of exploring a new region of
observational parameter space, it is the study of compact objects such
as white dwarfs, neutron stars and stellar-mass black holes which
benefits the most from high-speed observations. This is because the
dynamical timescales of compact objects range from seconds in white
dwarfs to milliseconds in neutron stars and black holes, which means
that the rotation and pulsation of these objects or material in close
proximity to them (e.g. in an accretion disc) tends to occur on
timescales of milliseconds to seconds. Other areas of astrophysics
which benefit from observations obtained on these timescales are
studies of eclipses, transits and occultations, where increased
time-resolution can be used to (indirectly) give an increased spatial
resolution.

Studying the above science requires a photometer with the following
capabilities:

\begin{enumerate}

\item {\em Short exposure times (from milliseconds to seconds).} There
  is little point in going much faster than milliseconds, as the
  gravity of compact objects does not allow for bulk motions on
  timescales below this. There are exceptions to this rule, however;
  for example, microsecond time-resolution would allow the study of
  magnetic instabilities in accreting systems, and nanosecond
  time-resolution would open up the field of quantum optics (see
  \citealt{dravins94}).

\item {\em Negligible dead time between exposures.} It takes a finite
amount of time to archive a data frame and start the next exposure. To
preserve time resolution, it is essential that this dead time is a
small fraction of the exposure time.

\item {\em Multi-channels (3 or more) covering a wide wavelength
range.}  Ideally, one would obtain spectra at high temporal resolution
in order to fully characterise the source of variability (e.g. to
determine its temperature).  The faintness of most compact objects
precludes spectroscopy, unfortunately. At a minimum, therefore, at least 
three different pass-bands covering as wide a portion of the optical 
spectrum as possible need to be observed, as this would allow a blackbody 
spectrum to be distinguished from a stellar spectrum. It is particularly
important that one of the three channels is sensitive to the far blue
(i.e. between approximately 3000--4000\AA), as the flickering and
oscillations observed in many accreting binaries is much more
prominent at these wavelengths (see, for example, \citealt{marsh98}).

\item {\em Simultaneous measurement of the different wavelength
    bands.} The requirement for multi-channel photometry must be tied
  to a requirement that each channel is observed simultaneously. A
  single-channel instrument with a filter wheel, for example, could
  obtain data in more than one colour by changing filters between each
  exposure or obtaining a full cycle in one filter and then a second
  cycle in another filter. The first technique results in poor time
  resolution and both the first and second techniques are
  observationally inefficient (relative to an instrument which can
  record multi-channel data simultaneously). In addition, the
  non-simultaneity of both techniques makes them unsuitable for the
  study of colour variations which occur on timescales shorter than
  the duration of observations in a single filter.

\item {\em Imaging capability.} A photometer with an imaging
capability is essential to observe variable sources. This is because,
unless the atmospheric conditions are perfectly photometric, it is
necessary to simultaneously measure the target, comparison stars and
sky background so that any variability observed can be unambiguously
assigned to the correct source. Although non-imaging photometers based
on multiple photomultiplier tubes can work around this, they still
suffer from their dependency on a fixed aperture which is larger than
the seeing disc and which therefore increases the sky noise
substantially. This is not a problem with imaging photometers, in
which the signal-to-noise of the extracted object counts can be
maximised using optimal photometry techniques (e.g. \citealt{naylor98}).

\item {\em High efficiency and portability.} General studies of
variability, i.e. the study of any object which eclipses, transits,
occults, flickers, flares, pulsates, oscillates, erupts, outbursts
or explodes, involves observing objects which span a huge range in
brightness. The faintest (e.g. the pulsars) are extremely dim and
observing them at high temporal resolution is a photon-starved
application on even the largest aperture telescopes currently
available. Others (e.g.  cataclysmic variables) are relatively
bright and can be observed at sub-second time-resolution at
excellent signal-to-noise on only a 2-m class telescope. Any 
instrument designed specifically for high-speed observations must
therefore be portable, so that an appropriate aperture telescope can
be selected for the project at hand, and highly efficient at
recording incident photons in order to combat the short exposure times
required.

\end{enumerate}

\section{Design}

With the exception of ULTRACAM, none of the instruments listed in
table~\ref{tab:competitors} meet all six of the scientific
requirements given above. In this section, we present an outline of
ULTRACAM's design.

\subsection{Optics}

The starting point for the ULTRACAM optical design was the required
field of view. The field of view has to be large enough to give a
significant probability of finding a comparison star of comparable
brightness to our brightest targets. The probability of finding a
comparison star of a given magnitude depends on the search radius and
the galactic latitude of the star. Using the star counts listed by
\cite{simons95}, the probability of finding a star of magnitude $R=12$
at a galactic latitude of 30$^{\circ}$ (the all-sky average) is 80\%
if the search radius is 5 arcminutes.\footnote{An on-line comparison
  star probability calculator can be found at
  http://www.shef.ac.uk/physics/people/vdhillon/ultracam.}  Most of
the target types discussed in section~\ref{sec:requirements} are
significantly fainter than $R=12$, so a field of view of 5 arcminutes
virtually guarantees the presence of a suitable comparison
star.\footnote{The study of transitting extrasolar planets is an
  exception to this rule. The brightest of these have magnitudes of
  $R\sim8$ and require fields of view of approximately 30 arcminutes
  to provide an 80\% chance of finding a comparison star of similar
  brightness.}

The next consideration in the ULTRACAM optical design was the required
pixel scale. Arguably the best compromise in terms of maximising
spatial resolution whilst minimising the contributions of readout
noise and intra-pixel quantum efficiency variations
(e.g. \citealt{jorden93}) is to use a pixel scale which optimally samples
the median seeing at the telescope. On the WHT, the median seeing
ranges from 0.55 to 0.73 arcseconds depending on the method of
measurement \citep{wilson99}, so a pixel scale of 0.3 arcseconds
provides a good compromise. The E2V 47-20 CCDs used in ULTRACAM (see
section~\ref{sec:detectors}) were selected because their 1024 pixels
on a side delivers a field of view of 5 arcminutes at a scale of 0.3
arcseconds/pixel, matching the requirements listed above perfectly.

There are a total of 30 optical elements in ULTRACAM, all of which
have a broad-band anti-reflection coating with transmission shown in
figure~\ref{fig:filters}. The maximum number of optical elements
encountered by a photon passing through ULTRACAM is 14 (see
figure~\ref{fig:raytrace}).  Light from the telescope is first
collimated and then split into three beams by two dichroic
beamsplitters. Each beam is then re-imaged by a camera onto a
detector, passing through a filter and CCD window along the way. Each
of these components is discussed in turn below.

\subsubsection{Collimator}

The collimator feeds collimated light to the re-imaging cameras,
partially corrects for the aberrations in the telescope it is to be
mounted on, and produces an image of the telescope pupil sufficiently
remote to accommodate the dichroics between the last collimator
element and the cameras. As the first optical element, the collimator
also has to have a high transmission in the scientifically important
$\sim$3000--4000\AA\ region (see section~\ref{sec:requirements})
without introducing significant chromatic aberration. The chosen
glasses were N-PSK3, CaF$_2$ and LLF1.

The collimator shown in figure~\ref{fig:raytrace} is used when
mounting ULTRACAM on the f/11 Cassegrain focus of the WHT. It can be
replaced with another collimator designed to accommodate the different
optical characteristics of other telescopes. We already have a second
collimator for use with the f/8 Cassegrain focus of the 2.3-m
Aristarchos Telescope in Greece, and a third for use at the f/15
Nasmyth focus of the VLT. Each of these collimators have identical
mounting plates on their lens barrels, so it is a simple matter to
switch between them on the opto-mechanical chassis
(section~\ref{sec:mechanics}). Note that the rest of the ULTRACAM
optics remain unchanged when moving between different telescopes.

\subsubsection{Dichroics}

Two dichroic beamsplitters divide the light from the collimator into
three different beams, which shall hereafter be referred to as the
`red', `green' and `blue' channels. Each dichroic consists of a
UV-grade fused silica substrate with a long-wave pass (LWP) coating
that reflects incident light with wavelengths shorter than the
cut-point whilst transmitting longer wavelengths. Such LWP dichroics
generally have a higher throughput than SWP dichroics and their
throughput has been further enhanced by coating the back surface of
each dichroic with a broad-band anti-reflection coating. The ULTRACAM
dichroics were manufactured by CVI Technical Optics, Isle of Man.

\begin{figure*}
\centering
\includegraphics[width=12cm]{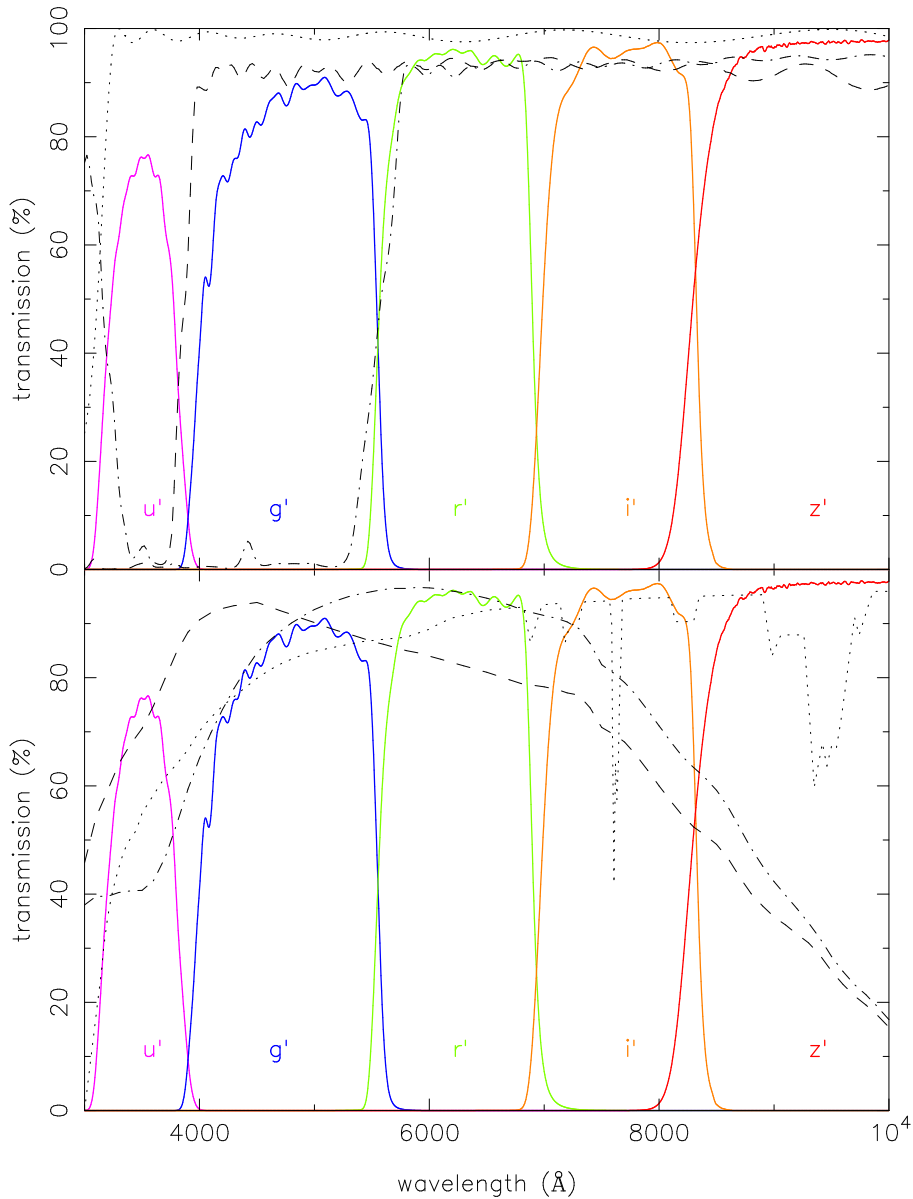}
\caption{Top: Transmission profiles of the ULTRACAM SDSS filter-set
  (solid lines), the anti-reflection coating used on the ULTRACAM
  lenses (dotted line), and the two dichroics (dashed line and
  dashed-dotted line). Bottom: Transmission profiles of the ULTRACAM
  SDSS filter-set (solid lines) and the atmosphere for unit airmass
  (dotted line). Also shown are the quantum efficiency curves of the
  blue and green CCDs (dashed line), and the red CCD (dashed-dotted
  line).}
\label{fig:filters}
\end{figure*}

To keep the cost of the custom optics as low as possible, whilst still
retaining maximum throughput and image quality, it was decided to
dedicate the blue channel of ULTRACAM to the scientifically-essential
SDSS $u'$-band. As a result, the 50\% cut point of the first dichroic
was set to 3870\AA\ (see figure~\ref{fig:filters}). A similar argument
forced us to dedicate the green channel of ULTRACAM to the SDSS
$g'$-band, giving a 50\% cut point for the second dichroic of 5580\AA.
In both cases, the cut-points are measured using incident light which
is randomly polarised.

It should be noted that the ULTRACAM dichroics operate in a collimated
beam. This has the great advantage that ghosts produced by reflections
off the back surface of the dichroics are in focus (but slightly
aberrated) and fall more-or-less on top of the primary image. This
means that the light in the ghosts, which is typically less than
0.001\% of the primary image, is included in the light of the target
when performing photometry, thereby maximising the signal-to-noise of
the observations and removing a potential source of systematic error.

\subsubsection{Re-imaging cameras}

\begin{figure*}
\centering
\includegraphics[width=15cm]{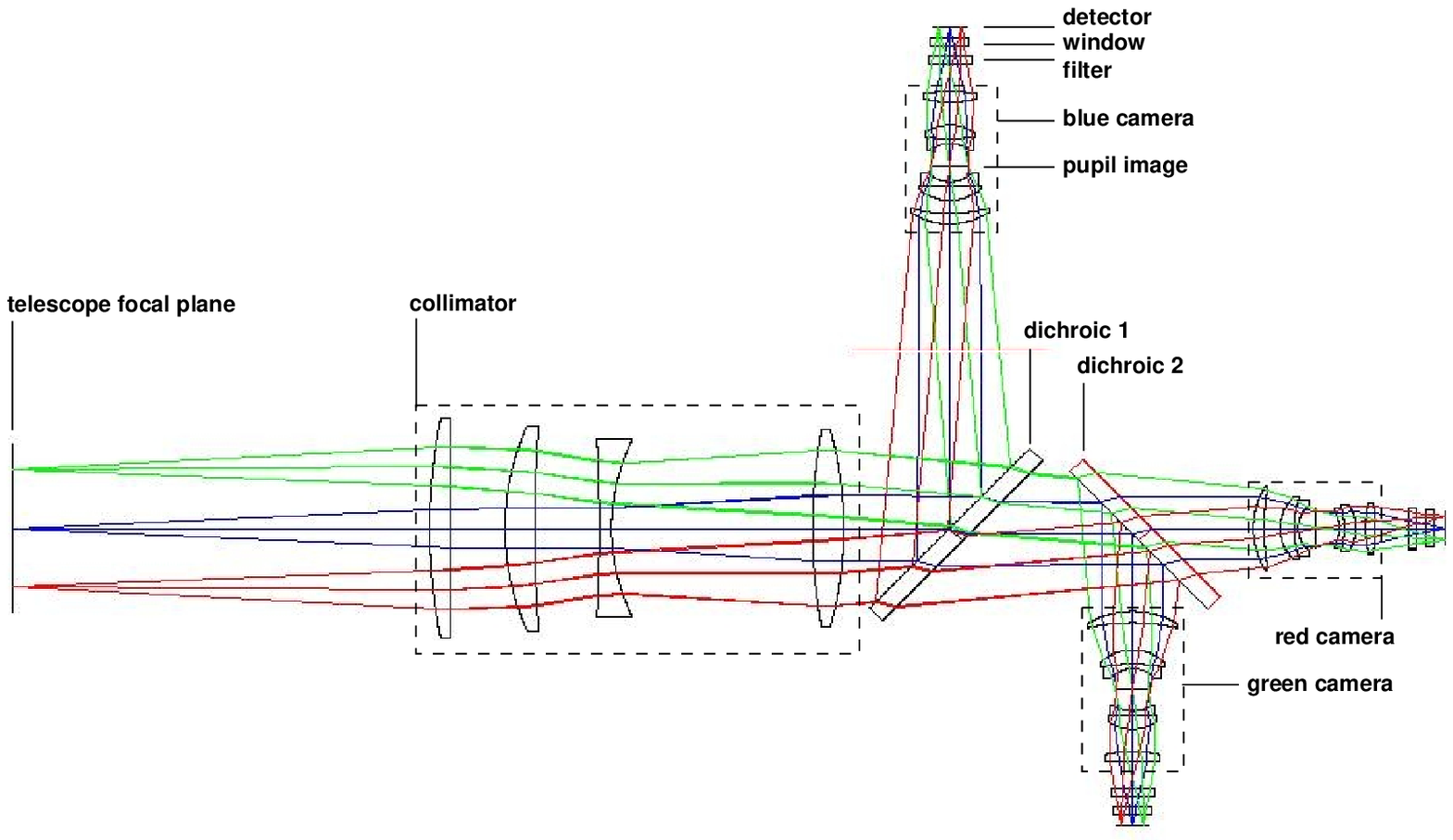}
\caption{Ray-trace through the ULTRACAM optics, showing the major optical
components: the collimator, dichroics, cameras, filters and detector windows.
The diagram is to scale -- the largest lens is in the collimator and has a 
diameter of 120 mm.}
\label{fig:raytrace}
\end{figure*}

The three re-imaging cameras are based upon a four-element
double-Gauss type lens, with two glass types per camera to minimise
costs (N-SK16 and LLF1 in the blue camera, N-LAK10 and SF2 in the
green camera, N-LAK22 and N-SF1 in the red camera). The glass
thickness has also been kept to a minimum, particularly in the blue
arm, to keep the throughput as high as possible. The cameras have been
optimised for an infinite object distance, with a field angle dictated
by the required field size. The re-imaged pupil is located between
the first and second pairs of elements (shown by the horizontal/vertical bars
near the centre of each camera in figure~\ref{fig:raytrace}), where the 
camera stop would be positioned. The back focal length of the cameras is
sufficient to allow space for the filter and CCD window. The cameras and 
collimators used in ULTRACAM were manufactured by Specac Ltd., UK.

\subsubsection{Filters and windows}

The Sloan Digital Sky Survey (SDSS) photometric system
\citep{fukugita96} was adopted as the primary filter set for
ULTRACAM. This system is becoming increasingly prevalent in
astrophysics and consists of five colour bands ($u^{\prime}$,
$g^{\prime}$, $r^{\prime}$, $i^{\prime}$ and $z^{\prime}$) that divide
the entire range from the atmospheric cut-off at $\sim 3000$\AA\ to
the sensitivity limit of CCDs at $\sim 11\,000$\AA\ into five
essentially non-overlapping pass-bands. The fact that SDSS filters
show only negligible overlap between their pass-bands is one of the
main reasons they have been adopted for use in ULTRACAM, as the
dichroic cut-points then have only a negligible effect on the shape of
the filter response.  The alternative Johnson-Morgan-Cousins system
($U$, $B$, $V$, $R_C$, $I_C$) suffers from overlapping pass-bands
which would be substantially altered when used in conjunction with the
ULTRACAM dichroics. The thinned CCDs used in ULTRACAM would also
suffer from fringing when used with $R_C$ due to the extended red tail
of this filter's bandpass, but this is eliminated by the sharp red
cut-off in SDSS $r^{\prime}$. The SDSS filters used in ULTRACAM were
procured from Asahi Spectra Ltd., Tokyo.

As well as SDSS filters and clear filters (for maximum throughput in
each channel), ULTRACAM also has a growing set of narrow-band filters,
including CIII/NIII+HeII (central wavelength, $\lambda_{c}=4662$\AA;
FWHM, $\Delta\lambda=108$\AA), NaI ($\lambda_{c}=5912$\AA,
$\Delta\lambda=312$\AA), H$\alpha$ ($\lambda_{c}=6560$\AA,
$\Delta\lambda=100$\AA), red continuum ($\lambda_{c}=6010$\AA,
$\Delta\lambda=118$\AA) and blue continuum ($\lambda_{c}=5150$\AA,
$\Delta\lambda=200$\AA).  All ULTRACAM filters are 50$\times$50 mm$^2$
and approximately 5 mm thick, but have been designed to have identical
optical thicknesses so that their differing refractive indices are
compensated by slightly different thicknesses, making the filters
interchangeable without having to significantly refocus the
instrument. Clearly, it is only possible to use these filters in
combinations compatible with the cut points of the two dichroics, such
as $u^{\prime}g^{\prime}r^{\prime}$, $u^{\prime}g^{\prime}i^{\prime}$,
$u^{\prime}g^{\prime}z^{\prime}$, $u^{\prime}g^{\prime}$ clear,
$u^{\prime}$ CIII/NIII+HeII red-continuum, $u^{\prime}g^{\prime}$
NaI, or $u^{\prime}$ blue-continuum H$\alpha$.

The final optical element encountered by a photon in ULTRACAM is the
CCD window. This allows light to fall on the chip whilst retaining the
vacuum seal of the CCD head (see section~\ref{sec:detectors}). The
windows are plane, parallel discs made of UV grade fused silica and
coated with the same anti-reflection coating as the lenses (see
figure~\ref{fig:filters}).

\subsection{Detectors}
\label{sec:detectors}

ULTRACAM uses three E2V 47-20 CCDs as its detectors. These are frame
transfer chips with imaging areas of 1024$\times$1024 pixels$^2$ and
storage areas of 1024$\times$1033 pixels$^2$, each pixel of which is
13 $\mu$m on a side. To improve quantum efficiency, the chips are
thinned, back-illuminated and anti-reflection coated with E2V's
enhanced broad-band astronomy coating (in the case of the blue and
green chips) and standard mid-band coating (in the case of the red
chip) -- see figure~\ref{fig:filters}.  The chips have a single serial
register which is split into two halves, thereby doubling the frame
rate, and each of these two channels has a very low noise amplifier at
its end, delivering readout noise of only $\sim 3.5\,e^-$ at pixel
rates of 11.2 $\mu$s/pixel and $\sim 5\,e^-$ at 5.6 $\mu$s/pixel.
The devices used in ULTRACAM are grade 0 devices, i.e. they are of the
highest cosmetic quality available. The full well capacity of these
devices is $\sim$ 100\,000$e^-$, and the devices are hence operated at
approximately unity gain with the 16-bit analogue-to-digital
converters in the CCD controller (see section~\ref{sec:das}).

In order for dark current to be a negligible noise source, it must be
significantly lower than the number of photons received from the
sky. In the worst case, in which an ULTRACAM observation is being
performed in the $u^{\prime}$-band when the Moon is new, on a dark
site (such as La Palma) and on a small aperture telescope (such as a
1-m), the number of photons from the sky incident on ULTRACAM would be
only $\sim 0.3\,e^-$/pixel/s. In this most pessimistic scenario,
therefore, the dark current must remain below $\sim 0.1\,e^-$/pixel/s
to be a negligible noise source. Fortunately, the E2V 47-20 CCDs can
be run in inversion mode (e.g. \citealt{mclean97}), which at a chip
temperature of 233~K delivers a dark current of only $\sim
0.05\,e^-$/pixel/s, beating our requirement by a factor of two. This
is a relatively high temperature, so the chips can be cooled by a
three-stage thermo-electric cooler (TEC) utilising the Peltier
effect. To maintain cooling stability, and hence dark current
stability, the hot side of the TEC is itself maintained at a constant
temperature of 283~K using a recirculating water chiller. Note that
this chiller is also used to cool the CCD controller (see
section~\ref{sec:das}), which is located at the base of the instrument
and can become extremely hot during operation.

The great advantage of using thermo-electric cooling as opposed to liquid
nitrogen, for example, is that the resulting CCD heads are very small
and lightweight.  With three such heads in ULTRACAM, this has
dramatically decreased both the mass and volume of the instrument. To
reduce conductive heating and to prevent condensation on the CCDs
whilst observing, the heads have been designed to hold a vacuum of
below $10^{-3}$\,Torr for several weeks, and we also blow dry air or
nitrogen gas across the front of the head to prevent condensation from
forming on the CCD window.

\subsection{Mechanics}
\label{sec:mechanics}

Small-aperture telescopes generally have much smaller size and mass
constraints on instruments than large-aperture telescopes. Given the
requirement for portability detailed in
section~\ref{sec:requirements}, we therefore designed the mechanical
structure of ULTRACAM assuming the space envelope and mass limits of a
typical 2-m telescope (e.g. Aristarchos).

The mechanical structure of ULTRACAM is described in detail by
\cite{stevenson04}. Briefly, it is required to: i) provide a stable
platform on which to mount the optics, CCD heads and CCD controller;
ii) allow easy access to the optics and CCD heads for alignment,
filter changes and vacuum pumping; iii) provide alignment mechanisms
for the optics and CCD heads; iv) exhibit low thermal expansion, as
all three cameras must retain parfocality; v) exhibit low flexure
(less than 1 pixel, i.e. 13 $\mu$m) at any orientation, so stars do
not drift out of the small windows defined on the three chips; vi) be
electrically and thermally isolated from the telescope in order to
reduce pickup noise and prevent the water chiller from attempting to
cool the entire telescope structure.

\begin{figure*}
\centering
\includegraphics[width=8cm]{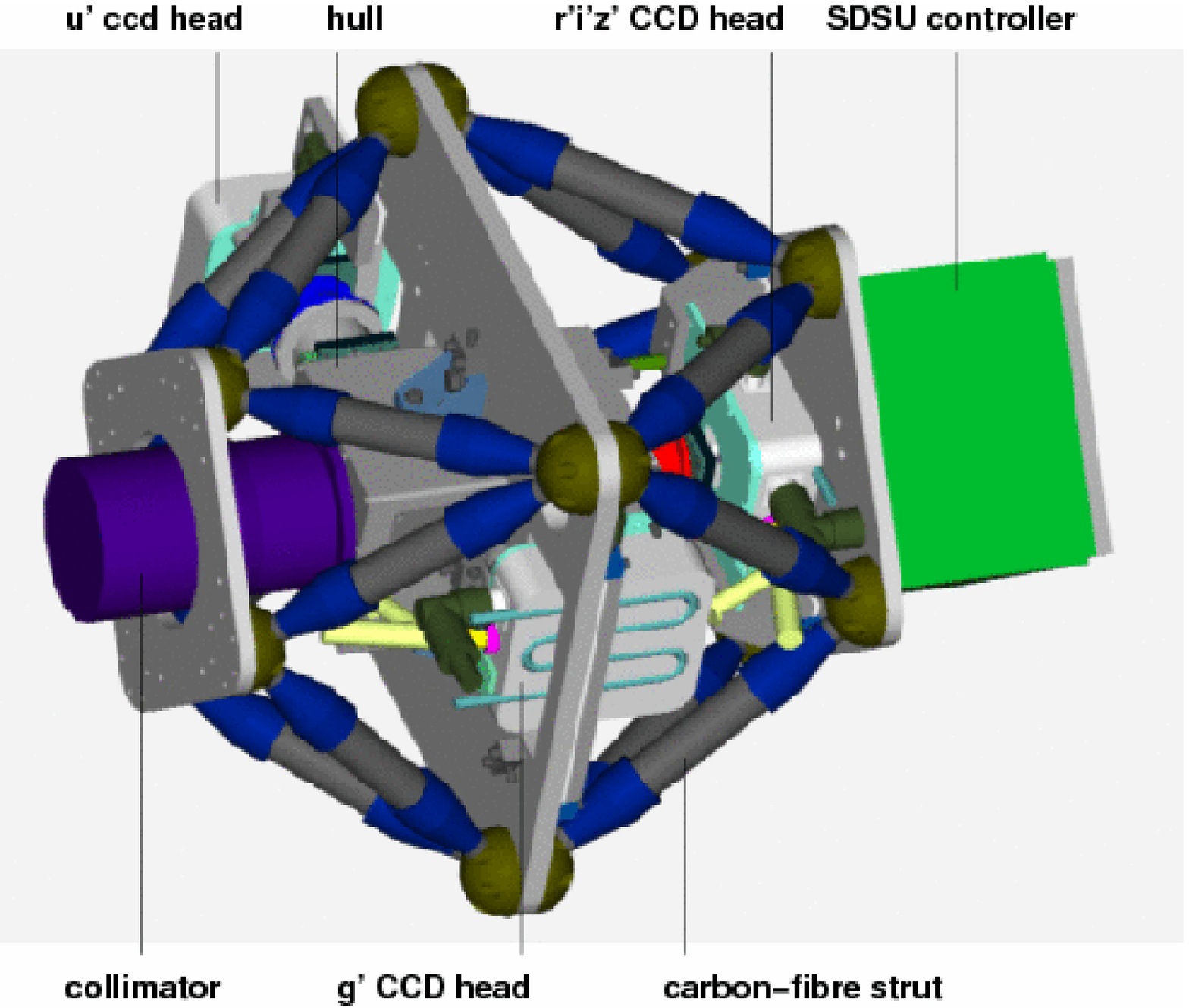}\hspace{0.5cm}\includegraphics[width=8cm]{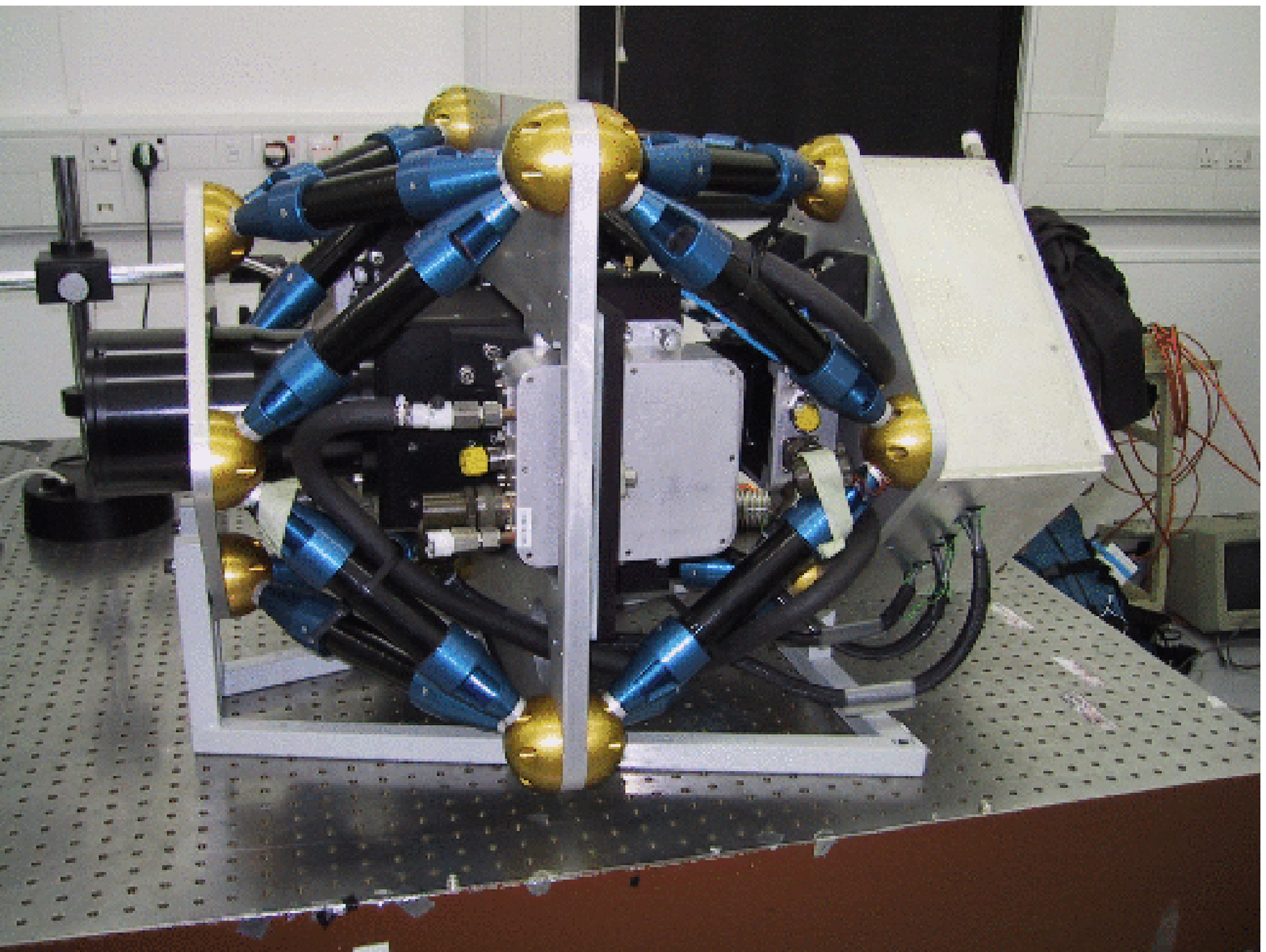}

\vspace*{0.5cm}

\includegraphics[width=8cm]{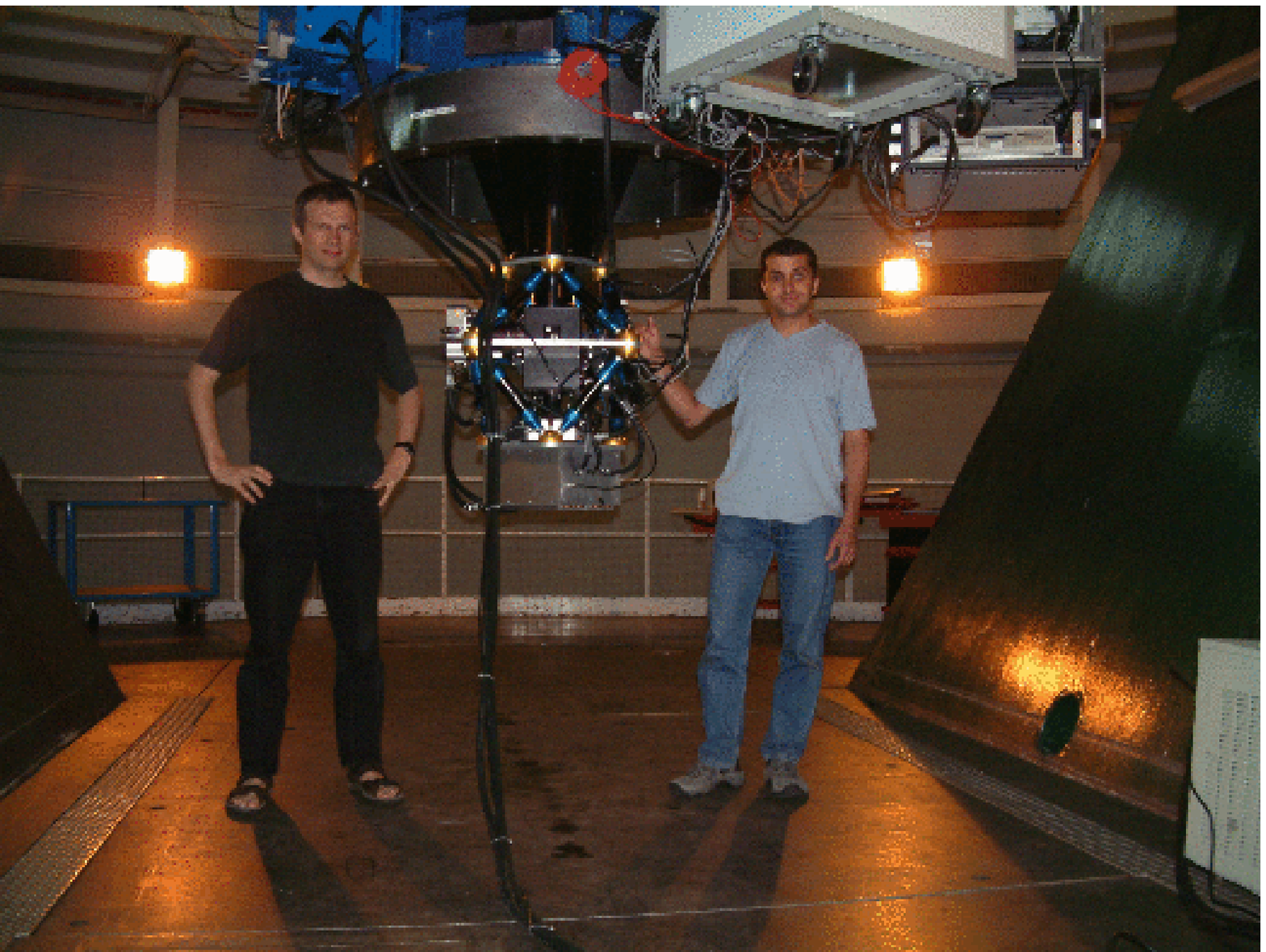}\hspace{0.5cm}\includegraphics[width=8cm]{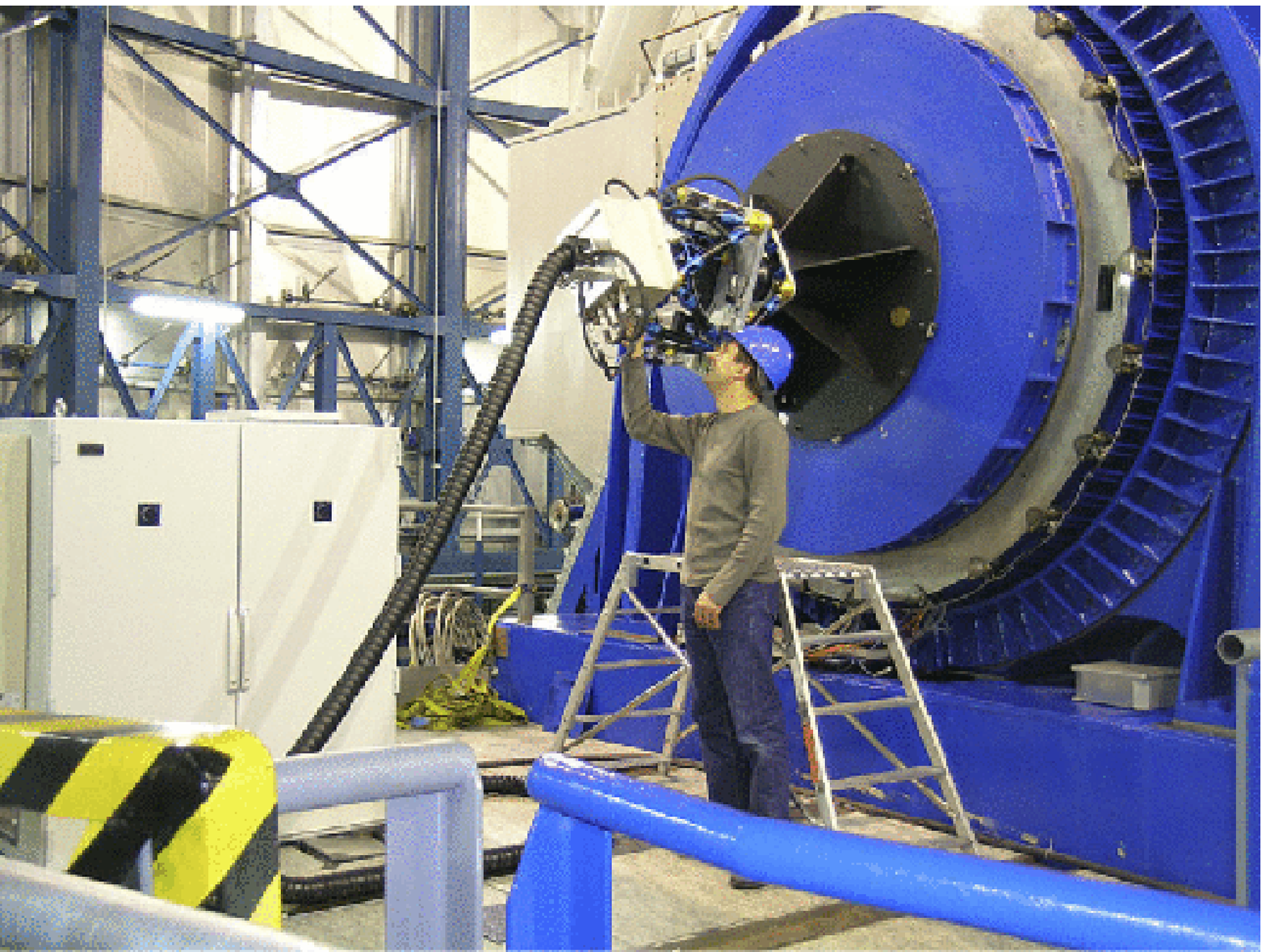}

\caption{Top left: CAD image of the ULTRACAM opto-mechanical chassis,
  highlighting some of the components described in the text. Top
  right: Photograph of ULTRACAM in the test focal station of the WHT
  (courtesy Sue Worswick). Bottom left: Photograph of ULTRACAM mounted
  on the Cassegrain focus of the WHT. Bottom right: Photograph of
  ULTRACAM mounted on the Visitor Focus of the VLT (Nasmyth on
  Melipal).}
\label{fig:mech}
\end{figure*}

The mechanical structure chosen for ULTRACAM is a double octopod, as
shown in figure~\ref{fig:mech}, which provides a rigid and open
framework meeting all of the requirements described above. The
Serrurier trusses of the double octopod are made of carbon fibre,
which offers similar structural strength to steel but at a five-fold
reduction in mass and a low thermal expansion coefficient. All of the
remaining parts of the mechanical chassis are of aluminium, giving a
total mass for the instrument of only 82 kg and an overall length of
just 792 mm (including the CCD controller and WHT collimator).

ULTRACAM requires a mounting collar to interface it to a telescope
focus.  The collar, which is constructed of steel, places ULTRACAM at
the correct distance from the telescope focal plane. A layer of G10/40
isolation material is placed between the collar and the top plate of
ULTRACAM to provide thermal and electrical isolation from the
telescope. The mounting collar also contains a motorised focal-plane
mask. This is an aluminium blade which can be moved in the focal plane
to prevent light from falling on regions of the CCD chip outside the
user-defined windows typically used for observing. Without this mask,
the light from bright stars falling on the active area of the chip
above the CCD windows would cause vertical streaks in the windows (see
figure 1 of \cite{dhillon05} for an example). The mask also prevents
photons from the sky from contaminating the background in drift-mode
windows (see section~\ref{sec:modes} and \cite{stevenson04} for
details).

\subsection{Data acquisition system}
\label{sec:das}

In this section we provide a brief overview of the ULTRACAM data
acquisition system. A much more detailed description can be found in
the paper by \cite{beard02}.

\subsubsection{Hardware}

\begin{figure*}
\centering
\includegraphics[width=10cm]{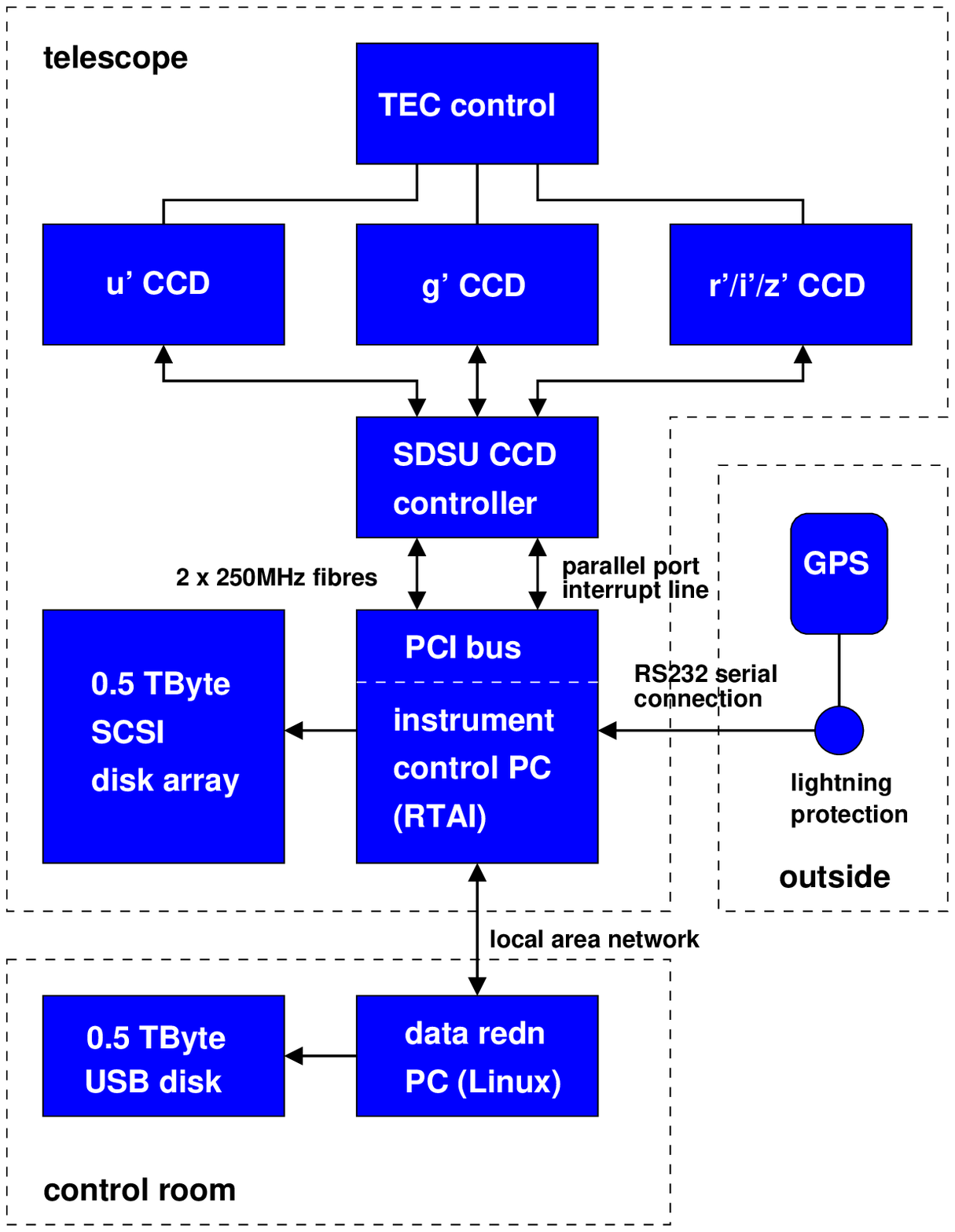}
\caption{Schematic showing the principal hardware components of the
ULTRACAM data acquisition system.  The connections between the
hardware components and their locations at the telescope are also
indicated.}
\label{fig:das}
\end{figure*}

Figure~\ref{fig:das} shows the data acquisition hardware used in
ULTRACAM. Data from the three CCD chips are read out by a San Diego
State University (SDSU) Generation III CCD controller
(\citealt{leach98}, \citealt{leach00}). The SDSU controller, which is
in wide use at many of the world's major ground-based telescopes, was
adopted by the ULTRACAM project due to its user programmability, fast
readout, low noise and ability to operate several multiple-channel
CCDs simultaneously. The SDSU controller is hosted by a rack-mounted
dual-processor PC running Linux patched with RealTime Application
Interface (RTAI) extensions. RTAI is a real-time version of Linux
which enables a user to have complete control over when a particular
task is carried out by a processor, irrespective of what else might be
running within the multi-tasking environment. For example, in a normal
Linux system undergoing heavy input/output activity (e.g. due to the
running of a data reduction task on a large file on disk), the delay
between requesting a command to be performed and its execution can be
hundreds of milliseconds in the worst case. With RTAI, this delay is
reduced to below $\sim 10\,\mu$s. In ULTRACAM, RTAI is used to provide
strict control over one of the two processors, so as to obtain
accurate timestamps from the Global Positioning System (GPS) antenna
located outside the dome and connected to the PC via a serial port
(see section~\ref{sec:gps} for details).

The instrument control PC communicates with the SDSU controller via a
Peripheral Component Interconnect (PCI) card and two 250 MHz optical
fibres. As well as communicating through the fibres, the SDSU
controller also has the ability to interrupt the PC using its parallel
port interrupt line, which is required to perform accurate
timestamping (see section~\ref{sec:gps}). Data from the CCDs are passed
from the SDSU PCI card to the PC memory by Direct Memory Access (DMA),
from where the data is written to a high-capacity SCSI disk array. All
of the real work in reading out the CCDs is performed by the SDSU
controller; the PCI card merely forwards commands and data between the
instrument control PC and the SDSU controller.

\subsubsection{Software}

The SDSU controller and PCI card both have on-board digital signal
processors (DSPs) which can be programmed by downloading assembler
code from the instrument control PC. An ULTRACAM user wishing to take
a sequence of windowed images, for example, would load the relevant
DSP application onto the SDSU controller (to control the CCDs) and PCI
card (to handle the data). The user can also modify certain
parameters, such as the exposure time or binning factors (see
section~\ref{sec:modes} for a full list), by writing the new values
directly to the DSP's memory.

All communication within the ULTRACAM system, including the loading of
DSP applications on the SDSU controller, is via Extensible Markup
Language (XML) documents transmitted using the Hyper Text Transfer
Protocol (HTTP) protocol. This is an international communications
standard, making the ULTRACAM data acquisition system highly portable
and enabling users to operate the instrument using any interface able
to send XML documents via HTTP (e.g. a web browser, perl scripts,
java).

\subsection{Timestamping}
\label{sec:gps}

Stamping CCD frames with start times accurate to a small fraction of
the typical exposure time is a key requirement for any astronomical
instrument. In ULTRACAM, with frame rates of up to 500 Hz, this
requirement is particularly severe, as it demands timestamping
accurate to a fraction of a millisecond. Without this level of
accuracy, for example, it would be impossible to determine the rate of
decrease in the orbital periods of interacting binary stars as
measured from eclipse timings \citep{brinkworth06}, or compare the
pulse arrival times in the optical and X-ray light from Anomalous
X-ray Pulsars \citep{dhillon05}.

Whenever an exposure is started, the SDSU controller sends an
interrupt to the instrument control PC which, thanks to the use of
RTAI, {\em instantaneously} (i.e. within $\sim 10\,\mu$s) writes the
current time to a First-In First-Out (FIFO) buffer.  A description of
how the current time is determined is given below.  The data handling
software then reads the timestamp from the FIFO and writes it to the
header of the next buffer of raw data written to the PC memory. In
this way, the timestamps and raw data always remain
synchronised. Moreover, as the SDSU controller reads out all three
chips simultaneously, the red, green and blue channels of ULTRACAM
also remain perfectly synchronised.

The clock provided on most PC motherboards is not able to provide the
current time with sufficient accuracy for our purposes, as it
typically drifts by milliseconds per second. The solution adopted
by ULTRACAM is to use a PCI-CTR05 9513-based counter/timer board,
manufactured by Measurement Computing Corporation, in conjunction with
a Trimble Acutime 2000 smart GPS antenna. Every 10 seconds, the GPS
antenna reports UTC to an accuracy of 50 nanoseconds. At the same
time, the number of ticks reported by the counter board is
recorded. At a later instant, when a timestamp is requested, the
system records the new value reported by the counter board, calculates the
number of ticks that have passed since the last GPS update, multiplies
this by the duration of a tick (which we have accurately measured in
the laboratory) and adds the resulting interval to the previous UTC
value reported by the GPS. Since the counter board ticks at 1 MHz, to
an accuracy of 100ppm, this is a much more reliable method of
timestamping than using the system clock of the PC. 

Laboratory measurements indicate that the timestamping in ULTRACAM has
a relative (i.e. frame-to-frame) accuracy of $\sim 50\,\mu$s. The
absolute timing accuracy of ULTRACAM has been verified to an accuracy
of $\sim 1$ millisecond by comparing observations of the Crab pulsar
with the ephemeris of \cite{lyne05} [see \cite{stevenson04} for details].

\subsection{Pipeline data reduction system}

ULTRACAM can generate up to 1 MByte of data per second. In the course
of a typical night, therefore, it is possible to accumulate up to 50
GBytes of data, and up to 0.5 TBytes of data in the course of a
typical observing run. To handle these high data rates, ULTRACAM has a
dedicated pipeline data reduction system\footnote{Available for
  download at http://deneb.astro.warwick.ac.uk/\newline
  phsaap/software/ultracam/html/index.html.}, written in C++, which
runs on a Linux PC or Mac located in the telescope control room and
connected to the instrument control PC via a dedicated 100BaseT local
area network (see figure~\ref{fig:das}).

Data from a run on an object with ULTRACAM is stored in two files, one
an XML file containing a description of the data format, and the other
a single, large unformatted binary file containing all the raw data and
timestamps. This latter file may contain millions of individual data
frames, each with its own timestamp, from each of the three ULTRACAM
CCDs. The data reduction pipeline grabs these frames from the SCSI
disk array by sending HTTP requests to a file server running on the
instrument control PC.

The ULTRACAM data reduction pipeline has been designed to serve two
apparently conflicting purposes. Whilst observing, it acts as a
quick-look data reduction facility, with the ability to display images
and generate light curves in real time, even when running at the
highest data rates of up to 1 MByte per second and at the highest
frame rates of up to 500 Hz. After observing, the pipeline acts as a
fully-featured photometry reduction package, including optimal
extraction \citep{naylor98}. To enable quick-look reduction whilst
observing, the pipeline keeps many of its parameters hidden to the
user and allows the few remaining parameters to be quickly skipped
over to generate images and light curves in as short a time as
possible. Conversely, when carefully reducing the data after a run,
every single parameter can be tweaked in order to maximise the
signal-to-noise of the final data.

\subsection{Readout modes}
\label{sec:modes}

\begin{figure*}
\centering
\includegraphics[width=14cm]{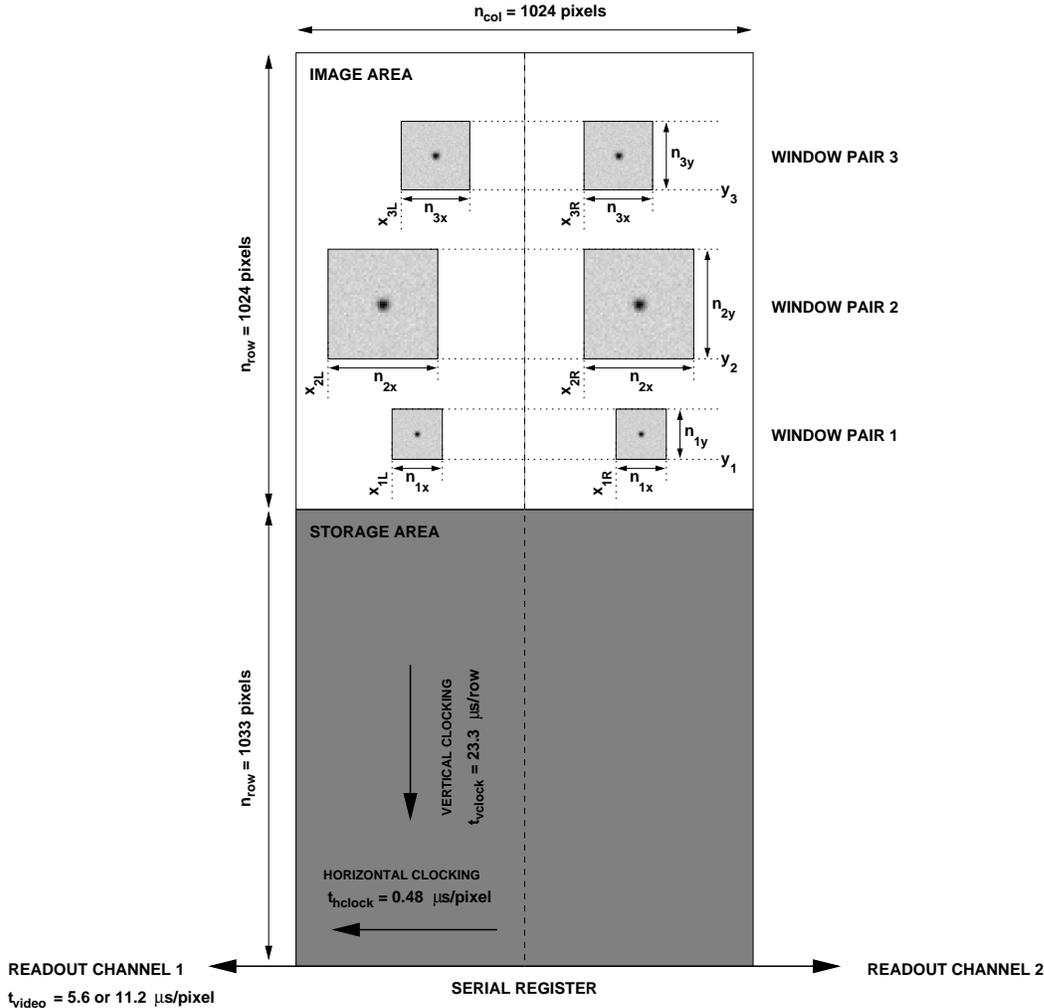}
\caption {Pictorial representation of an ULTRACAM CCD. The location of
  the image area, storage area and serial register are shown, along
  with the vertical and horizontal clocking directions. It can be seen
  that there are two readout channels, separated by the vertical
  dashed line, with the left-hand half of the chip read out via
  channel 1, and the right-hand half via channel 2. In two-window
  mode, only window pair 1 is present. In four-window mode, window
  pairs 1 and 2 are present, and in six-window mode all three window
  pairs are present. The chip can also be read out in full-frame mode,
  in which case the entire image area is selected (see
  figure~\protect\ref{fig:crab}). The window parameters and pixel
  rates defined in appendix~\ref{appendix} are also shown.}
\label{fig:frame}
\end{figure*}

Figure~\ref{fig:frame} can be used to illustrate how the
frame-transfer chips in ULTRACAM are able to deliver high frame
rates. A typical observation with ULTRACAM consists of the observation
of the target star in one user-defined CCD window and a nearby
non-variable comparison star in another, i.e. window pair 1 in
figure~\ref{fig:frame}. Once the exposure is complete, the image area
is shifted into the storage area, a process known as {\em vertical
  clocking}. This is very rapid, taking only 23.3\,$\mu$s per row and
hence $\sim$24 milliseconds to shift the entire 1024 rows into the
storage area. As soon as the image area is shifted in this way, the
next exposure begins.  Whilst exposing, the previous image in the
storage area is shifted onto the serial register and then undergoes
{\em horizontal clocking} to one of two readout ports, where it is
digitised. In other words, the previous frame is being read out whilst
the next frame is exposing, thereby reducing the dead time to the time
it takes to shift the image into the storage area, i.e. 24
milliseconds.

ULTRACAM has no shutter -- the fast shifting of data from the image
area to the storage area acts like an electronic shutter, and is far
faster than conventional mechanical shutters. This does cause some
problems, however, such as vertical trails of star-light from bright
stars, but these can be overcome in some situations by the use of a
focal-plane mask (see section~\ref{sec:mechanics}).

Setting an exposure time with ULTRACAM is a more difficult concept
than in a conventional non-frame-transfer camera. This is because
ULTRACAM attempts to frame as fast as it possibly can, i.e. it will
shift the image area into the storage area as soon as there is room in
the storage area to do so. Hence, the fastest exposure time is given
by the fastest time it takes to clear sufficient room in the storage
area, which in turn depends on the number, location, size and binning
factors of the windows in the image area, as well as the vertical
clocking and digitisation times, all of which are variables in the
ULTRACAM data acquisition system. To obtain an arbitrarily long
exposure time with ULTRACAM, therefore, an {\em exposure delay} is
added prior to the vertical clocking to allow photons to accumulate in
the image area for the required amount of time. Conversely, to obtain
an arbitrarily short exposure time with ULTRACAM, it is necessary to
set the exposure delay to zero and adjust the window, binning and
digitisation parameters so that the system can frame at the required
rate. As it takes 24 ms to vertically clock the entire image area into
the storage area, this provides a hard limit to the maximum frame rate
of $\sim 40$~Hz, with a duty cycle (given by the exposure time divided
by the sum of the exposure and dead times) of less than 3\%. Adopting
a more acceptable duty cycle of 75\% results in a useable limit of only
$\sim 10$ Hz.

Clearly, an alternative readout strategy is required in order to push
beyond the $\sim 10$ Hz frame-rate barrier and approach the desired
kilo-Hertz frame rates required to study the most rapid
variability. For this purpose, we have developed {\em drift mode},
which is described in detail in Appendix~\ref{appendix}. Briefly, in
drift mode the windows are positioned on the border between the image
and storage areas and, instead of vertically clocking the entire image
area into the storage area, only the window is clocked into the (top
of) the storage area. A number of such windows are hence present in
the storage area at any one time. This dramatically reduces the dead
time, as now the frame rate is limited to the time it takes to clock
only a small window into the storage area.  For example, in the case
of two windows of size 24$\times$24 pixels$^2$ and binned 4$\times$4,
it is possible to achieve a frame rate of $\sim 500$ Hz with a duty
cycle of $\sim 75$\%. This is currently the highest frame rate that
ULTRACAM has achieved on-sky whilst observing a science target. It is
worth noting that at these high frame rates it is the speed at which
charge can be shifted along the serial register on the ULTRACAM CCDs
(currently 0.48 $\mu$s/pixel), rather than the digitisation time, that
limits the frame rate. With larger windows and hence lower frame
rates, the reverse is true.

\begin{figure*}
\centering
\includegraphics[width=5.5cm,angle=270]{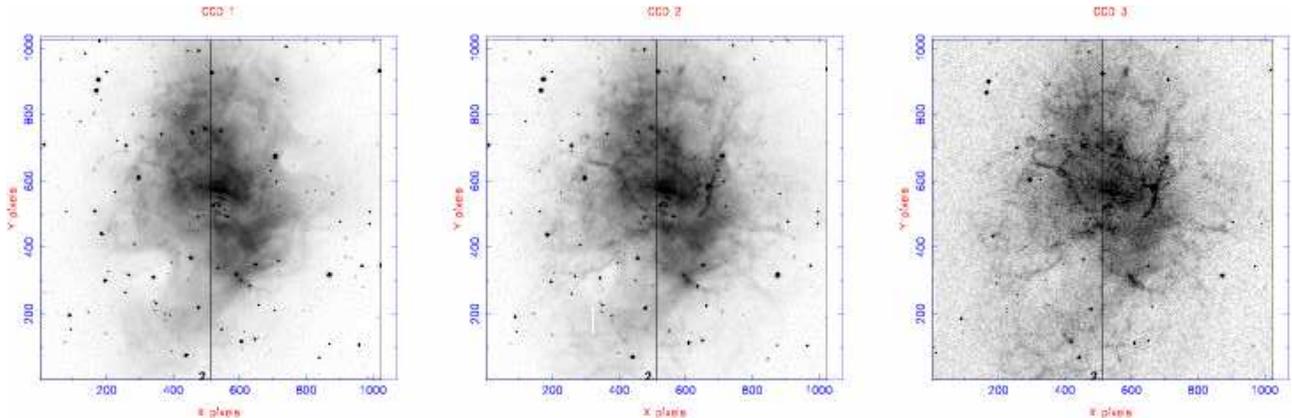}
\caption {Full-frame ULTRACAM image of the Crab Nebula (M1) in, from
  left-to-right, $i^{\prime}$, $g^{\prime}$ and $u^{\prime}$.  The
  image was taken using the WHT on 2002 September 11 and represents a
  total exposure time in each filter of 413 seconds. The Crab Pulsar
  is at pixel position (540, 525), just to the right of centre.  The
  resulting light curve of the Crab Pulsar was used to calibrate the
  ULTRACAM GPS timestamping system \protect\citep{stevenson04}. The
  field of view is 5 arcminutes, with North approximately at the
  top. The vertical line represents the border between the two readout
  channels on each chip, and the blank rows/columns at the top/right of
  the image indicate the underscan/overscan regions.}
\label{fig:crab}
\end{figure*}

Drift mode only offers the possibility of 2 windows and should only be
used when frame rates in excess of 10 Hz are required. This is because
the windows in drift mode spend longer on the chip and hence
accumulate more dark current and, without the use of the focal plane
mask, more sky photons. For frame rates of less than 10 Hz, ULTRACAM
offers normal two-window, four-window, six-window and full-frame readout
modes (the latter reading out in approximately 3 seconds with only 24
milliseconds dead time). An example of a full-frame image taken with
ULTRACAM is shown in figure~\ref{fig:crab}.

For some applications, e.g. when taking flat fields or observing
bright standard stars, it is desirable to use a full frame or two
large windows and yet have short exposure times. This is not possible
with the modes described above. To enable exposures times of
arbitrarily short length, therefore, ULTRACAM also offers the so-called
{\em clear} modes. These modes, which are available in full-frame and
two-window formats, clear the chip prior to exposing for the requested
amount of time. This means that any charge which has built up in the
image area whilst the previous exposure is reading out is discarded.
The disadvantage of this mode is that the duty cycle becomes very poor
(3\% in the case of a 0.1 second exposure time in full-frame-clear
mode).

A more detailed description of ULTRACAM's readout modes can be found
in appendix~\ref{appendix}

\section{Performance}

During the design phase of the project, a set of functional and
performance requirements for ULTRACAM were established against which
the instrument was tested during the commissioning phase. A detailed
description of how ULTRACAM performed when measured against these
requirements is given by \cite{stevenson04}. Other measures of the
performance of an astronomical instrument include how well the project
was managed, how reliable the instrument is in operation, how much
telescope time it wins, and how much science it enables. In what
follows, we attempt to summarise the performance of ULTRACAM measured
against these (very different) metrics.

\begin{table*}
\centering

\caption{Breakdown of the percentage of time spent observing  
  different classes of astronomical object with ULTRACAM on the 
  WHT and VLT. The right-hand column provides references to 
  some of the ULTRACAM papers published in each area. The entry
  for cataclysmic variables includes generic studies of accretion
  discs around white dwarfs (6.6\%) and pulsating white dwarfs in
  binaries (3.3\%). The entry for X-ray binaries is composed of systems with 
  black-hole primaries (7.9\%) and neutron-star primaries (9.5\%).}

\begin{tabular}{lrl}
\hline\noalign{\smallskip}
Target & Time & References \\
\hline\noalign{\smallskip}

Cataclysmic variables & 25.3\% & \cite{littlefair06c}, \cite{feline04} \\

X-ray binaries & 17.4\% & \cite{munoz-darias06}, \cite{shahbaz05} \\

sdB stars/asteroseismology & 11.2\% & \cite{aerts06}, \cite{jeffery04}\\

Eclipsing, detached white-dwarf/red-dwarf binaries & 10.4\% & \cite{brinkworth06}, \cite{maxted04} \\

Occultations by Titan, Pluto and Kuiper Belt Objects & 10\% & \cite{fitzsimmons07}, \cite{roques06} \\

Ultra-compact binaries & 6.6\% & \cite{barros07}, \cite{marsh06} \\

Pulsars & 5\% & \cite{dhillon06}, \cite{dhillon05} \\

Extrasolar planet transits & 3.7\% & \\

Flare stars & 3.3\% & \cite{mathioudakis06} \\

Isolated brown dwarfs & 2.9\% & \cite{littlefair06b} \\

Isolated white dwarfs & 1.7\% & \cite{silvotti06} \\

Gamma-ray bursts & 1.7\% & \cite{vreeswijk05} \\

Active galactic nuclei & 0.8\% & \\

\hline\noalign{\smallskip}
\end{tabular}
\label{tab:stats}
\end{table*}

\begin{enumerate}

\item ULTRACAM was delivered on budget ($\sim\pounds$300\,000) and 3
  months ahead of the three-year schedule set to design, build and
  commission it.

\item ULTRACAM is an extremely reliable instrument, as it has no
  moving parts. To date, ULTRACAM has suffered from $<2$\% technical
  downtime, and this has nearly all been due to unavoidable hardware
  failures (such as hard disks and power supplies).

\item ULTRACAM has met nearly all of its performance requirements,
  including throughput (50\% in $g^{\prime}$), image quality (seeing
  limited at 0.6"/0.3" on WHT/VLT), flexure (10 $\mu$m), timing
  accuracy (50 $\mu$s) and detector-limited data throughput. Using
  high-quality optics, detectors and coatings, ULTRACAM has achieved
  photometric zero points on the VLT of approximately
  $u^{\prime}=26.0$, $g^{\prime}=28.1$, $r^{\prime}=27.6$,
  $i^{\prime}=27.7$ and $z^{\prime}=26.7$.

\item The frame rates achieved on sky have ranged from 0.05 Hz to 500
  Hz on targets ranging in brightness from 8--26 magnitude. It can be
  seen that the maximum frame rate is a factor of two below the
  required 1 kHz value set at the start of the project. This is not a
  serious drawback, however, as the science performed by ULTRACAM to
  date (see table~\ref{tab:stats}) has almost all concentrated on the
  0.1--1 second regime.

\item The ULTRACAM project team has adopted a policy of open access,
  offering the instrument to any astronomer on a shared-risks,
  collaborative basis. As a result, ULTRACAM has been awarded a great
  deal of competitively allocated observing time: 89 nights on the WHT
  and 39 nights on the VLT in the 5 years since first light. A
  breakdown of how these 128 nights have been spent on different types
  of object is given in table~\ref{tab:stats}.

\item To date, a total of $\sim 25$ refereed papers have been
  published which are wholly or partly dependent on ULTRACAM
  data. Some examples of the research done with ULTRACAM are given in
  table~\ref{tab:stats}.

\item Any instrument which is not continually enhanced and upgraded
  will eventually cease to be competitive. For this reason, we have
  always pursued a vigorous enhancements programme with ULTRACAM,
  focussed on the three most important areas for high time-resolution
  astrophysics: increasing the maximum frame rate, reducing the
  readout noise and minimizing downtime/inefficiency. The performance
  of the instrument has improved by at least a factor of two in these
  three areas in the 5 years since first light.

\end{enumerate}

\section{Conclusions}

We have described the scientific requirements, design and performance
of ULTRACAM, and listed examples of the varied science that has been
performed with the instrument. As for the future, we intend to
continue to operate ULTRACAM for approximately one run per year on
both the VLT and the WHT.\footnote{Anyone interested in using ULTRACAM
  on a shared-risks, collaborative basis are encouraged to contact one
  of the first two authors.} We have also recently commissioned a
spectroscopic version of ULTRACAM, known as ULTRASPEC
\citep{dhillon07}. This instrument uses an electron-multiplying,
frame-transfer CCD and the ULTRACAM data acquisition system to deliver
high-speed, zero readout-noise spectroscopy with the EFOSC2 spectrograph
on the ESO 3.6-m telescope at La Silla. In the longer term, we are
exploring the possibility of building ULTRACAM-II, which would be
dedicated for use in Chile and which would have four channels, a
larger field of view and, if available, use multi-channel, split
frame-transfer electron-multiplying CCDs.

\section*{Acknowledgments}

There are many people who have contributed to the success of ULTRACAM
but who have not been included as co-authors on this paper. We would
like to thank all of them for their dedication to the project,
particularly the staff of the Central Mechanical Workshops at the
University of Sheffield, the UK Astronomy Technology Centre in
Edinburgh, the Isaac Newton Group of Telescopes on La Palma and the
European Southern Observatory at Paranal. We would also like to thank
PPARC for providing the funding to build, operate and exploit
ULTRACAM.

We dedicate this paper to the memory of our friend and colleague,
Emilios Harlaftis, whose enthusiasm for the project and desire to use
ULTRACAM on the new 2.3-m Aristarchos Telescope in Greece
significantly influenced the design of the instrument.

\bibliographystyle{mn2e}
\bibliography{abbrev,refs}

\appendix

\section{Modelling CCD frame rates}
\label{appendix}

In this section we expand on the brief description of ULTRACAM's
readout modes given in section~\ref{sec:modes} and provide an
algorithm to compute ULTRACAM exposure times and dead times (and hence
frame rates). The algorithm can straightforwardly be adapted to
predict the performance of any frame transfer CCD\footnote{An on-line
  ULTRACAM frame-rate calculator can be found at
  http://www.shef.ac.uk/physics/people/vdhillon/ultracam.}.

Experiments show that the algorithm is accurate to a few per cent, due
mainly to uncertainties in the precise values of the input parameters
such as the vertical clocking time, horizontal clocking time and
digitisation time. Note that we ignore the negligible dead-time
induced by the time it takes to execute the DSP code controlling the
readout process, and the negligible influence of the various dark
reference columns and rows on the calculated frame rates. For
simplicity, we also ignore the negligible delay induced by the extra
one or two pixels which must be added or subtracted from some of the
equations given below to allow for the fact that, for example, the
number of pixels in a window is given by the difference between its
end and start pixel positions $+1$.

Referring to figure~\ref{fig:frame}, we define the following symbols:

\begin{itemize}
\item[] {\bf $n_{row}$}: The total number of rows in the image area of
  the CCD. In ULTRACAM, this figure is 1024. For simplicity, we assume
  that the storage area has the same number of rows; in reality, the
  storage area of the CCDs used in ULTRACAM have 9 more rows than the image
  area.
\item[] {\bf $n_{col}$}: The number of columns in the image area. The
storage area has the same number of columns. The number of pixels
in the serial register is also equal to this number. In ULTRACAM, this
figure is 1024. 
\item[] {\bf $n_{1x}$, $n_{1y}$, $n_{2x}$, $n_{2y}$, $n_{3x}$, $n_{3y}$}:
The $x$ and $y$ sizes of each window in pairs 1, 2, and 3, respectively.
\item[] {\bf $x_{1L}$, $x_{1R}$, $x_{2L}$, $x_{2R}$, $y_{3L}$, $y_{3R}$}: 
The $x$ position in the left (L) and right (R) readout channels of the
start of window pair 1, 2 and 3, respectively.
\item[] {\bf $y_1$, $y_2$, $y_3$}: The $y$ position of the bottom
of window pair 1, 2, and 3, respectively.
\item[] {\bf $b_x$, $b_y$}: The binning factors in the $x$ (horizontal) and
$y$ (vertical) directions.
\item[] {\bf $n_{win}$}: In drift mode, the number of windows in the stack in 
the storage area.
\item[] {\bf $t_{vclock}$}: The time taken to vertically clock one
  row. In ULTRACAM, this has the same value (23.3 $\mu$s/row) in both
  the image and storage areas, but this need not necessarily be the
  case.
\item[] {\bf $t_{hclock}$}: The time taken to horizontally clock one pixel
along the serial register. In ULTRACAM, this is currently limited to
0.48 $\mu$s/pixel.
\item[] {\bf $t_{video}$}: The time taken to determine the charge
  content of a pixel (via correlated double sampling) and perform the
  analogue-to-digital conversion. Two speeds are currently used in
  ULTRACAM: slow (11.2~$\mu$s/pixel) and fast (5.6 $\mu$s/pixel).
\item[] {\bf $t_{delay}$}: The exposure delay, a user-defined pause in
the readout process which increases the time spent accumulating
photons in the image area.
\item[] {\bf $t_{pipe}$}: An additional exposure delay used in drift
  mode which ensures that the exposure times and the gaps between
  exposures are uniform.
\item[] {\bf $t_{inv}$}: The inversion time. This is the time
taken for the chip to come out of inversion mode in preparation for
the readout process. In ULTRACAM, this takes 110 $\mu$s.
\item[] {\bf $t_{clear}$}: The time taken to clear the chip, i.e. to 
vertically clock the entire image and storage areas and dump the charge.
\item[] {\bf $t_{frame}$}: The frame-transfer time, i.e. the time taken to 
transfer the entire image area into the storage area. 
\item[] {\bf $t_{y1}$, $t_{y2}$, $t_{y3}$}: The time taken to
  vertically shift window pair 1, 2 and 3, respectively, in the
  storage area in order to place them adjacent to the serial register.
\item[] {\bf $t_{line}$, $t_{line1}$, $t_{line2}$, $t_{line3}$}: For
  full-frame, window pair 1, 2 and 3, respectively, the time taken to
  shift one row of the storage area into the serial register, shift it
  along the serial register, determine the charge content of each
  pixel and then perform the analogue-to-digital conversion.
\item[] {\bf $t_{cycle}$}: The cycle time, defined as the total amount of time
it takes to expose and then read out an image.
\item[]{\bf $t_{exp}$}: The exposure time, defined as the time spent 
accumulating photons during which no clocking is performed in the image
area.
\item[]{\bf $t_{dead}$}: The dead time, defined as the time
during which the CCD is not accumulating photons. 
\item[] {\bf $\nu_{frame}$}: The frame rate, defined as  $\nu_{frame}=1/t_{cycle}$.
\end{itemize}

\subsection{Full-frame mode}
\label{fullframemode}

ULTRACAM can be read out in full-frame mode with and without an
overscan/underscan and with and without clearing the chip prior to
each exposure. Figure~\ref{fig:crab} shows an example full-frame image
taken by ULTRACAM, in which the overscan/underscan regions can be seen
at the right-hand side and top of the image.  For the sake of
simplicity, we shall not consider the overscan/underscan case here;
this effectively adds an additional 8 rows and 56 columns to the chip.

We first consider the case where the chip is cleared prior to each
exposure. The time taken to clear the chip is given by the time it
takes to vertically clock both the image and storage areas:

\begin{equation}
t_{clear} =  2 n_{row} \times t_{vclock}.
\end{equation}
Next, move the image into the storage area:

\begin{equation}
t_{frame} = n_{row} \times t_{vclock}.
\end{equation}
The time it then takes to shift one row of the storage area into the
serial register, shift it along the entire serial register, determine
the charge content of each pixel and then digitise the data is given
by:

\begin{equation}
t_{line} = (b_y \times t_{vclock}) + (\frac{n_{col}}{2} \times t_{hclock})
+ (\frac{n_{col}}{2 b_x} \times t_{video}),
\label{eqn:tline}
\end{equation}
where the factors of 2 are due to the fact that the ULTRACAM CCDs are
two-channel devices. The total time taken to read out one exposure is
hence the sum of the following:

\begin{equation}
t_{cycle} = t_{clear} + t_{delay} + t_{inv} + t_{frame} + 
(\frac{n_{row}}{b_y} \times t_{line}).
\end{equation}
Because the chip is cleared prior to each exposure, any charge
accumulated in the image area during the previous exposure is lost.
The exposure time is then simply:

\begin{equation}
t_{exp} = t_{delay}.
\end{equation}
Hence it is possible to obtain arbitrarily short exposure times in clear
mode.

Instead, if the chip is not cleared prior to each exposure, the total
time taken to read out one exposure is:

\begin{equation}
t_{cycle} = t_{delay} + t_{inv} + t_{frame} + (\frac{n_{row}}{b_y} \times t_{line}),
\end{equation}
i.e. there is no $t_{clear}$ term (which makes only a negligible
difference to the cycle time). Since all charge that accumulates in
the image area whilst the previous exposure is reading out is
recorded, the exposure time is given by:

\begin{equation}
t_{exp} = t_{cycle} - t_{frame}.
\end{equation}
Hence the shortest possible exposure time obtainable in no-clear mode
is given by the time it takes to read the entire chip out. In
ULTRACAM, this can be up to 6 seconds (when not binning and using the
slowest value for $t_{video}$).

The dead times of both the clear and no-clear modes are:

\begin{equation}
t_{dead}=t_{cycle}-t_{exp}.
\end{equation}
In the case of the no-clear mode, this means that
$t_{dead}=t_{frame}$.  This is negligible, since $t_{frame}$ is only
24 milliseconds in ULTRACAM.  In the case of the clear mode, however,
$t_{dead}=t_{cycle}-t_{delay} \simeq \frac{n_{row}}{b_y} \times
t_{line}$, which means that the dead-time is dependent on the time it
takes to read the entire chip out (i.e. up to 6 seconds).

\subsection{Two-windowed mode}
\label{twowindowedmode}

ULTRACAM is most often used for point-source photometry, in which case
it is usually not necessary to read out the entire array. Instead,
windows around the target and comparison stars are defined, as shown
in figure~\ref{fig:frame}, substantially increasing the frame rate and
reducing the amount of data accumulated.

It is possible to define either two, four or six windows with
ULTRACAM, with one window of each pair occupying the left-hand side of
the chip and the other window occupying the right-hand
side. Additionally, for each pair, the two windows must have the same
sizes, vertical start positions and binning factors, and they must not
overlap with each other. These rules significantly simplify the data
acquisition software and yet still give full flexibility when
observing, as any two stars can be located in two windows simply by
adjusting the telescope position, the instrument rotator angle and the
horizontal start positions and sizes of the windows.

In two-windowed mode, it is possible to read out the chip both with
and without clearing. We first consider the case where the chip is
cleared prior to each exposure. The time taken to clear the chip is
given by the time it takes to vertically clock the image and storage
areas:

\begin{equation}
t_{clear} =  2 n_{row} \times t_{vclock}.
\end{equation}
Next, move the windows into the storage area:
 
\begin{equation}
t_{frame} = n_{row} \times t_{vclock}.
\end{equation}

At this stage, the windows have the same vertical position in the
storage area as they had in the image area. Hence the next step is
to vertically shift the windows to place them adjacent to the serial
register:

\begin{equation}
t_{y1} = y_1 \times t_{vclock}.
\end{equation}

To simplify the data acquisition software, and ensure true simultaneity
between the three ULTRACAM chips, each of the windows is forced to
have the same pixel position on the red, green and blue
CCDs. Furthermore, the data acquisition system expects data from the
left-hand channel of each chip to be processed at the same time as
data from the right-hand channel, which effectively means that a
window in the left-hand channel must lie the same number of pixels
from the centre-line of the chip as a window in the right-hand
channel. In practice, such a rule would be very restrictive during
target acquisition, so instead a {\em differential shift} is
performed. On entering the serial register, each row of the window
closest to the centre-line of the chip is shifted by $\left|x_{1L} -
  (n_{col}-x_{1R}-n_{1x})\right|$ pixels until it lies the same
distance from the centre-line as the window in the other channel.

The time it takes to shift one row of the storage area into the serial
register, perform the differential shift, horizontally shift the
window to the output, determine the charge content of each pixel and
then digitise the data is given by:

\begin{equation}
t_{line1} = (b_y \times t_{vclock}) + (n_{hclock1} \times t_{hclock})
+ (\frac{n_{1x}}{b_x} \times t_{video}).
\label{eqn:tline1}
\end{equation}
Note that the factor of 2 present in equation~\ref{eqn:tline} is
absent from the equation for $t_{line1}$ above due to the fact that
one window is read out by one of the two channels and the second
window of the pair is read out via the other channel. $n_{hclock1}$ is
the total number of horizontal clocks required and is given by the sum
of the differential shift, the window size, and the number of rows
between the output and whichever window is closest to it. Once the
serial register has been clocked $n_{hclock1}$ times, both windows
will have been read out. The charge remaining in the serial register
is then dumped, rather than clocked out. This takes only $8 \times
t_{hclock}$ and therefore significantly reduces the readout time in
the case where the windows are defined to lie close to the vertical
edges of the detector. Strictly speaking, therefore, one should also
add 8 to $n_{hclock1}$.

The total time taken to read out one exposure is hence the sum of the
following:

\begin{equation}
t_{cycle} = t_{clear} + t_{delay} + t_{inv} + t_{frame} + t_{y1} +
(\frac{n_{1y}}{b_y} \times t_{line1}).
\end{equation}
Because the chip is cleared prior to each exposure, any charge
accumulated in the windows in the image area during the previous
exposure is lost.  The exposure time is then simply:

\begin{equation}
t_{exp} = t_{delay}.
\end{equation}
Hence, it is possible to obtain arbitrarily short exposure times in
clear mode.

Instead, if the chip is not cleared prior to each exposure, the total
time taken to read out one exposure is:

\begin{equation}
t_{cycle} = t_{delay} + t_{inv} + t_{frame} + t_{y1} + 
(\frac{n_{1y}}{b_y} \times t_{line1}),
\end{equation}
i.e. there is no $t_{clear}$ term (which makes only a negligible
difference to the cycle time). Since all charge that accumulates in
the image area whilst the previous exposure is reading out is
recorded, the exposure time is given by:

\begin{equation}
t_{exp} = t_{cycle} - t_{frame}.
\end{equation}
Hence the shortest possible exposure time obtainable in no-clear mode
is given by the time it takes to read the windows out. This can be
reduced by using smaller windows, binning, and/or ensuring the windows
are positioned so as to minimise $n_{hclock1}$ and $t_{y1}$.

The dead times of both the clear and no-clear modes are:

\begin{equation}
t_{dead}=t_{cycle}-t_{exp}.
\end{equation}
In the case of the no-clear mode, this means that
$t_{dead}=t_{frame}$.  This is negligible, since $t_{frame}$ is only
24 milliseconds in ULTRACAM.  In the case of the clear mode, however,
$t_{dead}=t_{cycle}-t_{delay} \simeq \frac{n_{1y}}{b_y} \times
t_{line1}$, which means that the dead-time is dependent on the time it
takes to read the windows out. 

\subsection{Four and six-windowed mode}

The four and six-windowed modes are only currently available in
no-clear mode. For the sake of brevity, we shall only outline the
four-windowed mode here -- it is a simple matter to extend the
algorithm to six windows.

To read out, the windows are first moved into the storage area:

\begin{equation}
t_{frame} = n_{row} \times t_{vclock}.
\end{equation}

The next step is to vertically shift the first pair of windows to 
place them adjacent to the serial register:

\begin{equation}
t_{y1} = y_1 \times t_{vclock}.
\end{equation}

Once this pair has been read out (see below), the second pair of windows
must be vertically shifted to the serial register:

\begin{equation}
t_{y2} = (y_2 - y_1 - n_{1y}) \times t_{vclock}.
\end{equation}

The differential shift for the first pair of windows is given in
appendix~\ref{twowindowedmode}. In the same way, the differential shift
for the second pair of windows is given by $\left|x_{2L} -
  (n_{col}-x_{2R}-n_{2x})\right|$ pixels.

For the first window pair, the time it takes to shift one row of the
storage area into the serial register, perform the differential shift,
horizontally shift the window to the output, determine the charge
content of each pixel and then digitise the data is denoted by
$t_{line1}$ and is given by equation~\ref{eqn:tline1}. Similarly,
for the second window pair, $t_{line2}$ is given by:

\begin{equation}
t_{line2} = (b_y \times t_{vclock}) + (n_{hclock2} \times t_{hclock})
+ (\frac{n_{2x}}{b_x} \times t_{video}),
\end{equation}
where $n_{hclock2}$ is the total number of horizontal clocks required
for the second window pair and is given by the sum of the differential
shift, the window size, and the number of rows between the output and
whichever window of the second pair is closest to it.

The total time taken to read out one exposure is hence the sum of the
following:

\begin{eqnarray}
t_{cycle} = t_{delay} + t_{inv} + t_{frame} + t_{y1} + t_{y2} \nonumber \\
+ (\frac{n_{1y}}{b_y} \times t_{line1}) + (\frac{n_{2y}}{b_y} \times t_{line2}).
\end{eqnarray}

Since there is no clear, all of the charge that accumulates in
the image area whilst the previous exposure is reading out is
recorded. The exposure time is therefore given by:

\begin{equation}
t_{exp} = t_{cycle} - t_{frame}.
\end{equation}
Hence the shortest possible exposure time obtainable in no-clear mode
is given by the time it takes to read the windows out. This can be
reduced by using smaller windows, binning, and/or ensuring the windows
are positioned so as to minimise $n_{hclock1}$, $n_{hclock2}$,
$t_{y1}$ and $t_{y2}$.

The dead time is: 

\begin{equation}
t_{dead}=t_{cycle}-t_{exp},
\end{equation}
which means that $t_{dead}=t_{frame}$.  This is negligible, since
$t_{frame}$ is only 24 milliseconds in ULTRACAM.

\subsection{Drift mode}

\begin{figure*}
\centering
\includegraphics[width=15cm]{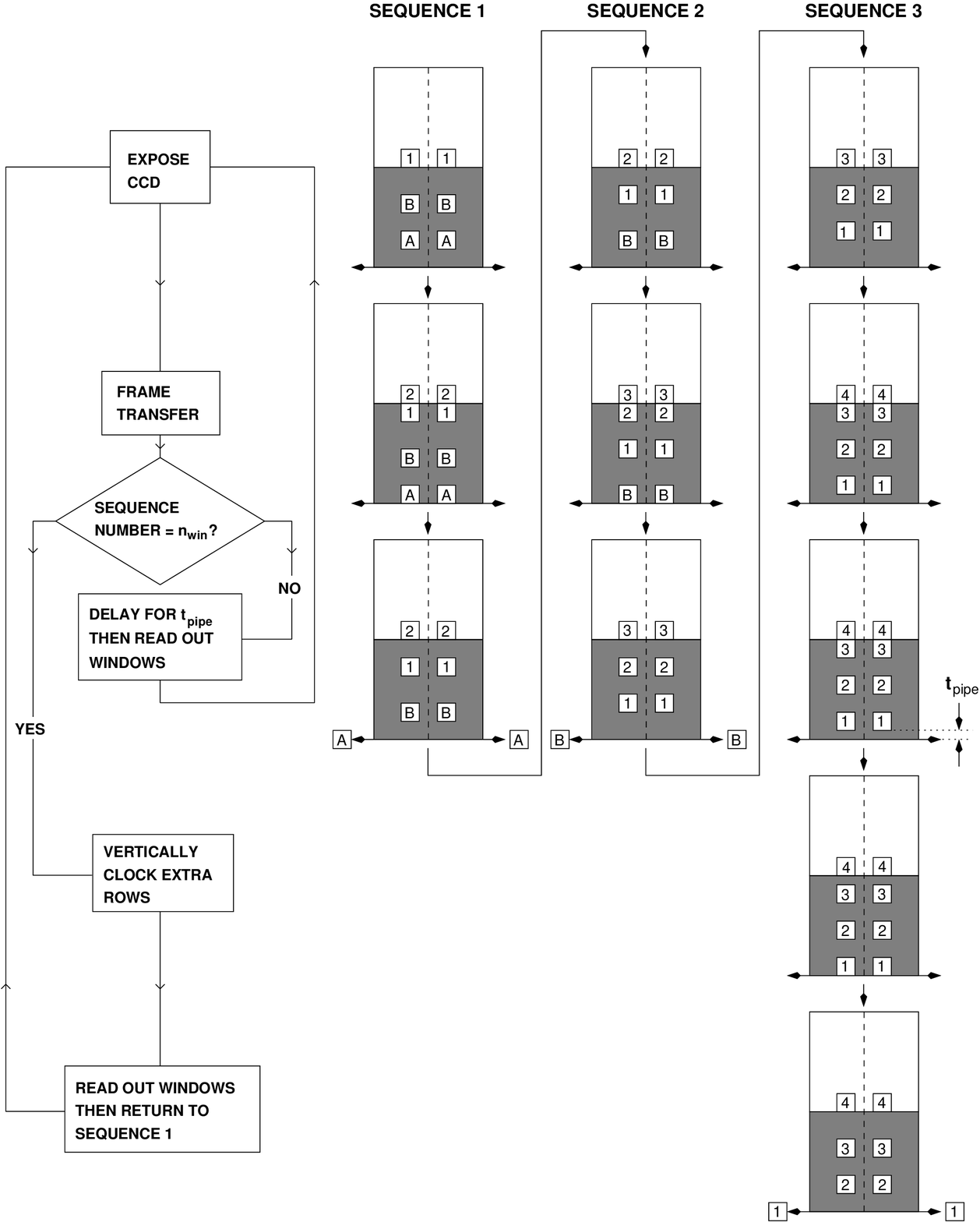}
\caption{Pictorial representation of the readout sequence in drift
  mode with $n_{win}=3$. Exposed windows form a vertical stack in the
  storage area. The storage area has 1033 rows, implying that the
  vertical gaps between the windows can never all be the same. To
  maintain uniform exposure times and intervals between exposures,
  therefore, a pipeline delay is added to sequences 1 and 2. On
  completion of sequence 3, the cycle begins again by returning to
  sequence 1.}
\label{fig:drift_mode}
\end{figure*}

Drift mode is the most complex of the ULTRACAM readout modes and is
used when frame rates in excess of $\sim 10$ Hz are required (see
section~\ref{sec:modes}). Due to its complexity, drift mode is only
available in two-windowed mode without clearing.

The readout sequence in drift mode is shown pictorially in
figure~\ref{fig:drift_mode}. It can be seen that the basic difference
between drift mode and the standard two-windowed mode is in the
frame-transfer process. In two-windowed mode, the entire image area is
clocked into the storage area, resulting in an irreducible minimum
dead time of 24 milliseconds. In drift mode, the two windows are
positioned on the border between the storage and image areas, and only
the rows containing the windows are clocked into the storage area on
completion of an exposure. Since each row takes only 23.3 $\mu$s to
vertically clock, a small window of, say, 24 rows takes only $\sim$0.5
milliseconds to move into the storage area, reducing the dead time by
almost a factor of 50.

Figure~\ref{fig:drift_mode} shows that, by moving the exposed windows
from the bottom of the image area into the top of the storage area, a
corresponding region at the bottom of the storage area must be moved
off the chip.  For this to occur rapidly, the charge must be dumped at
the serial register rather than clocked along it and digitised. Once
the dump is complete, a previously exposed window will lie at the
bottom of the storage area and will then be read out whilst another
window is exposing in the image area. In this way it can be seen that
a stack of exposed windows are accumulated in the storage area, with
each window separated by a gap equal to or greater than $n_{1y}$.

As shown below, the exposure time of a window at the bottom of the
image area is dependent on how long it takes to read out the window at
the bottom of the storage area. Unfortunately, the CCD used in
ULTRACAM has 1033 rows in its storage area, which is a prime
number. This means that one of the gaps between the windows in the
stack must be slightly larger than $n_{1y}$, and hence takes a
slightly longer time to read out. The resulting variable exposure
times and variable intervals between exposures would have a
detrimental effect on periodicity analyses, and hence a correction is
introduced. Specifically, a {\em pipeline delay}, $t_{pipe}$, is added
to each window bar the one with the larger gap, thereby equalising the
exposure times of each window in the stack and the time intervals
between them (see figure~\ref{fig:drift_mode}).

The number of windows, $n_{win}$, in the stack in the storage area
depends on the number of rows in the windows and the size of the
storage area, as follows:

\begin{equation}
n_{win} = \left\lfloor \left(\frac{n_{row}}{n_{1y}} + 1\right) / 2 \right\rfloor,
\label{eqn:nwin}
\end{equation}
where $n_{row} = 1033$ for the storage area of the ULTRACAM CCDs, and the
outer brackets on the right-hand side indicate that only the integer part of
the result is recorded. 

The pipeline delay can now be determined by calculating the extra number
of rows present in the largest gap between the windows in the stack:

\begin{equation}
t_{pipe} = (n_{row} - [(2 n_{win} - 1) \times n_{1y}]) \times t_{vclock}.
\end{equation}

With the above defined, we can now model the frame rate in drift mode.
On completion of a user-defined exposure delay, $t_{delay}$, the first
step in the drift-mode readout process is to move the window from the
image area to the top of the storage area:

\begin{equation}
t_{frame} = (n_{1y} + y_1) \times t_{vclock}.
\end{equation}

This will cause a previously exposed window close to the bottom of the
storage area to reach the serial register. The differential shift
required to ensure that the two windows lie the same distance from the
centre-line of the chip is given in section~\ref{twowindowedmode}.
The time it then takes to shift one row of the storage area into the
serial register, perform the differential shift, horizontally shift
the windows to the output, determine the charge content of each pixel
and then digitise is given by:

\begin{equation}
t_{line1} = (b_y \times t_{vclock}) + (n_{hclock1} \times t_{hclock})
+ (\frac{n_{1x}}{b_x} \times t_{video}),
\end{equation}
where $n_{hclock1}$ is the total number of horizontal clocks required
and is given by the sum of the differential shift, the window size,
and the number of rows between the output and whichever window is
closest to it.

The total time taken to read out one exposure is hence the sum of the
following:

\begin{equation}
  t_{cycle} = t_{delay} + t_{pipe} + t_{inv} + t_{frame} +
  (\frac{n_{1y}}{b_y} \times t_{line1}).
\end{equation}

Since there is no clear, the exposure time is given by:

\begin{equation}
t_{exp} = t_{cycle} - t_{frame}.
\end{equation}
Hence the shortest possible exposure time obtainable is given by the
time it takes to read the windows out. This can be reduced by using
smaller windows, binning, and ensuring the windows are positioned and
sized so as to minimise $n_{hclock1}$ and $t_{pipe}$. This latter fact
can be particularly important. For example, selecting windows of
$n_{1y}=24$ results in $n_{win}=22$ and $t_{pipe} =
t_{vclock}$. Selecting windows of $n_{1y}=23$, however, results in
$n_{win}=22$ but $t_{pipe} = 44 \times t_{vclock}$, i.e. an additional
exposure time of 1 millisecond.  In this way, it can be seen that
there are a series of ``special'' values of $n_{1y}$ in drift mode
which satisfy, for example, $t_{pipe} < 14 \times t_{vclock}$ (i.e.
an additional exposure time of $< 0.33$ milliseconds). These special
values are: $n_{1y}$ = 8, 10, 13, 18, 21, 24, 31, 38, 41, 49, 54, 60,
68, 79, 93, 114, 147, 206, 344.

The dead time is given by: 

\begin{equation}
t_{dead}=t_{cycle}-t_{exp},
\end{equation}
as for the rest of the time the windows are integrating on the sky.
This means that $t_{dead}=t_{frame}$. Since $t_{frame}$ depends on
$n_{1y}$ in drift mode, as opposed to $n_{row}$ in the other ULTRACAM
readout modes, the dead-time is minimised.

It should be noted that GPS timestamping in drift mode is slightly
more complicated than in the non-drift modes. As described in
section~\ref{sec:gps}, whenever an exposure is started the SDSU
controller sends an interrupt to the instrument control PC which
immediately writes the current time to the next frame of data received
by the PC. In the case of drift mode, however, the next frame of data
is not the most recently exposed window, but the window which happened
to be at the bottom of the stack in the storage area. There is hence a
loss of synchronization between timestamp and data frame. This is
easily rectified by the pipeline data reduction system, however, as it
is a simple matter to calculate which window each timestamp should be
attached to using equation~\ref{eqn:nwin}.

\label{lastpage}

\end{document}